\documentclass[journal]{IEEEtran}
\usepackage{amsmath,amsfonts}
\usepackage{algorithmic}
\usepackage{array}
\usepackage{subfig}
\usepackage{textcomp}
\usepackage{caption}
\usepackage{stfloats}
\usepackage{url}
\usepackage{verbatim}
\usepackage{graphicx}
\usepackage{cite}
\usepackage{amssymb}
\usepackage{bbm}
\usepackage{xcolor}
\usepackage{amsthm}
\usepackage{dsfont}
\usepackage[linesnumbered,ruled,vlined]{algorithm2e}
\begin{document}

\title{Cooperative Sensing and Heterogeneous Information Fusion in VCPS: A Multi-agent Deep Reinforcement Learning Approach}

\author{Xincao Xu, Kai Liu,~\IEEEmembership{Senior Member,~IEEE,} Penglin Dai,~\IEEEmembership{Member,~IEEE,} Ruitao Xie, \\ Jingjing Cao, and Jiangtao Luo,~\IEEEmembership{Senior Member,~IEEE}
\thanks{Manuscript received.}
\thanks{Xincao Xu and Kai Liu are with the College of Computer Science, Chongqing University, Chongqing 400040, China. (e-mail: near, liukai0807@cqu.edu.cn).}
\thanks{Penglin Dai is with the School of Computing and Artificial Intelligence, Southwest Jiaotong University, Chengdu 611756, China, and also with the National Engineering Laboratory of Integrated Transportation Big Data Application Technology, Chengdu 611756, China. (e-mail: penglindai@swjtu.edu.cn).}
\thanks{Ruitao Xie is with the College of Computer Science and Software Engineering, Shenzhen University, Shenzhen 518060, China. (e-mail: xie@szu.edu.cn).}
\thanks{Jingjing Cao is with the School of Transportation and Logistics Engineering, Wuhan University of Technology, Hubei 430063, China. (e-mail: bettycao@whut.edu.cn).}
\thanks{Jiangtao Luo is with the Electronic Information and Networking Research Institute, Chongqing University of Posts and Telecommunications, Chongqing 400065, China (e-mail: luojt@cqupt.edu.cn).}
}



\maketitle

\begin{abstract}
Cooperative sensing and heterogeneous information fusion are critical to realize vehicular cyber-physical systems (VCPSs). 
This paper makes the first attempt to quantitatively measure the quality of VCPS by designing a new metric called Age of View (AoV). 
Specifically, we first present the system architecture where heterogeneous information can be cooperatively sensed and uploaded via vehicle-to-infrastructure (V2I) communications in vehicular edge computing (VEC). 
Logical views are constructed by fusing the heterogeneous information at edge nodes. 
Further, we formulate the problem by deriving a cooperative sensing model based on the multi-class M/G/1 priority queue, and defining the AoV by modeling the timeliness, completeness and consistency of the logical views.  
On this basis, a multi-agent deep reinforcement learning solution is proposed. 
In particular, the system state includes vehicle sensed information, edge cached information and view requirements. 
The vehicle action space consists of the sensing frequencies and uploading priorities of information. 
A difference-reward-based credit assignment is designed to divide the system reward, which is defined as the VCPS quality, into the difference reward for vehicles.
Edge node allocates V2I bandwidth to vehicles based on predicted vehicle trajectories and view requirements.
Finally, we build the simulation model and give a comprehensive performance evaluation, which conclusively demonstrates the superiority of the proposed solution. 
\end{abstract}

\begin{IEEEkeywords}
Vehicular cyber-physical system, edge computing, cooperative sensing, heterogeneous information fusion, multi-agent deep reinforcement learning
\end{IEEEkeywords}

\section{Introduction}
Recent advances in sensing technologies and vehicular communications drive the development of vehicular cyber-physical systems (VCPSs) \cite{jia2015survey}, which is a key enabler of the next generation of intelligent transportation systems (ITSs).
In VCPS, heterogeneous information such as traffic light status, vehicle locations, point cloud data and surveillance videos can be cooperatively sensed and uploaded by vehicles.
The view, which is logical mapping of physical status of the elements in vehicular networks, such as the locations, velocities, and heading directions of vehicles, and the status of traffic lights can be constructed at the edge node based on the fusing of sensed information. 
The particular elements to be constructed in a logical view is determined by the specific ITS applications.
On the other hand, vehicular edge computing (VEC) \cite{liu2019hierarchical} becomes a promising paradigm in supporting high-density vehicular communications, massive data transmission, and adaptive computation offloading at the edge of vehicular networks.
Therefore, it is of great significance to investigate quality cyber-physical fusion in VEC.

Great efforts have been devoted to data dissemination \cite{liu2020fog}\cite{singh2020intent}, information caching \cite{zhang2021digital, dai2020deep, su2018edge} and task offloading \cite{shang2021deep}\cite{liao2020learning} in vehicular networks.
However, none of them have investigated the synergistic effects of cooperative sensing and heterogeneous information fusion.
A number of studies have been studied on predicting \cite{zhang2019cyber}\cite{zhang2020data}, scheduling \cite{li2020cyber}\cite{lian2020cyber}, and controlling \cite{dai2016convex, hu2017cyber, lv2018driving, xu2020vehicular} technologies in VCPS, which facilitated the implementation of various ITS applications.
Nevertheless, they are based on assumed sufficient and reliable information collected by the edge/cloud nodes.
Several studies have concerned the information quality evaluation in VCPS \cite{liu2014temporal, dai2019temporal, liu2014realtime, rager2017scalability, yoon2021performance}.
However, they only evaluated the quality at the data item level, while ignoring the quality evaluation for the fusing of heterogeneous information.
Some studies have focused on vehicle sensing and information fusion using deep reinforcement learning (DRL) in vehicular networks \cite{zhao2020social, dong2020spatio, mlika2022deep}, but they are not suitable in modeling multiple vehicle scenarios.
A few literatures have applied the multi-agent DRL into vehicular networks \cite{xu2023joint}\cite{he2021efficient}.
However, none of the solutions can be directly applied in VCPS for cooperative sensing and heterogeneous information fusion.
To the best of our knowledge, this is the first work on investigating the synergistic effect of cooperative sensing and heterogeneous information fusion by quantitatively measuring the quality of VCPS. 

The critical issues and challenges to be addressed in this paper are summarized as follows.
First, the physical information is highly dynamic.
Therefore, it is critical to consider the synergistic effect of the sensing frequency, queuing delay, and transmission delay to ensure information freshness and timeliness.
Second, the physical information is temporal-spatial correlated.
Meanwhile, vehicles have different sensing capacities and they are scheduled in a distributed manner.
Consequently, vehicles are expected to be cooperated in information sensing and uploading to reduce resource consumption and enhance information quality.
Third, vehicle-to-infrastructure (V2I) communications have limited radio coverage, and unreliable due to the nature of wireless communications.
Thus, it is also critical to alleviate the impact of intermittent connection and packet loss during uploading.
Fourth, the physical information is intrinsically heterogeneous in terms of distribution, updating frequency, and modality, which brings great challenges to the quality modeling of information fusion.

With above motivations, we jointly investigate the evaluation metric and the scheduling algorithm, aiming at enhancing the quality of VCPS by synergizing the cooperative sensing and heterogeneous information fusion.
The main contributions are outlined as follows. 
\begin{enumerate}
\item A novel problem is investigated in VCPS by integrating the sensing, uploading, modeling and evaluation of heterogeneous  information. In particular, a cooperative sensing model is derived based on the multi-class M/G/1 priority queue and the Shannon theory. On this basis, a new metric called Age of View (AoV) is designed to evaluate the timeliness, completeness, and consistency of heterogeneous information in VCPS. To the best of our knowledge, this is the first work on quantitatively evaluating the quality of VCPS with the consideration of unique characteristics captured by the newly designed metic AoV.
\item A dedicated solution is proposed based on multi-agent deep reinforcement learning. Specifically, vehicles act as independent agents with action space of sensing frequencies and uploading priorities. Then, a different reward (DR) based credit assignment scheme is designed to evaluate the contributions of individual vehicles on view construction, so as to enhance the evaluation accuracy in term of the action of each agent. Further, the solution manages to achieve smaller action space of each agent and speed up the convergency compared with conventional DRL algorithms. Meanwhile, a V2I bandwidth allocation (VBA) scheme is designed at the edge node based on vehicle trajectories and view requirements.
\item A comprehensive performance evaluation is conducted based on real-world vehicular trajectories. The proposed solution and four competitive algorithms, including random allocation (RA), centralized deep deterministic policy gradient (C-DDPG) \cite{mlika2022deep}, multi-agent actor-critic (MAC) \cite{he2021efficient} and MAC with  V2I bandwidth allocation scheme (MAC-VBA) are implemented. The simulation results demonstrated that the proposed solution outperforms RA, C-DDPG, MAC, and MAC-VBA by around 61.8\%, 23.8\%, 22.0\%, and 8.0\%, respectively, in terms of maximizing the VCPS quality, and speeds up the convergence by around 6.8$\times$, 1.4$\times$ and 1.3$\times$ compared with C-DDPG, MAC, and MAC-VBA, respectively.
\end{enumerate}

The rest of this paper is organized as follows.
Section II reviews the related work.
Section III presents the system architecture.
Section IV formulates the problem.
Section V proposes the solution. 
Section VI evaluates the performance.
Finally, Section VII concludes this paper and discusses future research directions.

\section{Related Work}

There have been numerous studies on data dissemination, information caching, and task offloading in vehicular networks.
Liu et al. \cite{liu2020fog} considered the cooperative data dissemination problem in a vehicular end-edge-cloud architecture, and proposed a clique searching-based scheduling scheme to enable collaborative data encoding and dissemination.
Singh et al. \cite{singh2020intent} proposed an intent-based network control framework, where a neural network is used to train the flow table and enables intelligent data dissemination.
Zhang et al. \cite{zhang2021digital} proposed a social-aware vehicular edge caching mechanism, which dynamically orchestrates the cache capability of edge nodes and smart vehicles according to user preference similarity and service availability.
Dai et al. \cite{dai2020deep} proposed a blockchain-enabled distributed information caching framework, which integrates DRL and permissioned blockchain and achieves intelligent and secure information caching.
Su et al. \cite{su2018edge} developed a dynamic information caching scheme based on the analyzed vehicular content request features.
Shang et al. \cite{shang2021deep} studied energy-efficient task offloading and developed a deep-learning-based algorithm to minimize the energy consumption.
Liao et al. \cite{liao2020learning} presented a task offloading strategy for air-ground integrated VEC that enables vehicles to learn long-term strategies with a multi-dimensional awareness of intent.
These studies mainly focused on scheduling algorithms for data dissemination, information caching, and task offloading in vehicular networks. However, none of them have investigated the synergistic effects of cooperative sensing and heterogeneous information fusion in VCPS.

Great efforts have been devoted to predicting, scheduling, and controlling technologies in VCPS.
Zhang et al. \cite{zhang2019cyber} proposed a hybrid velocity-profile prediction method, which integrates the traffic flow state with individual driving behaviors. 
Zhang et al. \cite{zhang2020data} predicted the vehicle status based on a lane-change behavioral prediction model and an acceleration prediction model.
Li et al. \cite{li2020cyber} considered vehicle mobility and developed a physical-ratio-K interference model-based broadcast scheme to ensure communication reliability.
Lian et al. \cite{lian2020cyber} presented a scheduling method for path planning based on an established map model to optimize the path utilization efficiency.
Dai et al. \cite{dai2016convex} proposed an autonomous intersection control mechanism to determine vehicle priorities for passing through intersections.
Hu et al. \cite{hu2017cyber} proposed a fuel-optimal controller to optimize the vehicle speed and continuously variable transmission gear ratio based on the leading vehicle status.
Lv et al. \cite{lv2018driving} presented an adaptive algorithm to control the vehicle acceleration under three typical driving styles with different protocol selections.
Xu et al. \cite{xu2020vehicular} proposed a vehicle collision warning scheme based on trajectory calibration by considering V2I communication delay and packet loss.
These studies focused on different technologies to support VCPS, such as trajectory predicting, path scheduling, and vehicle controlling, which facilitated the implementation of various ITS applications. Nevertheless, these studies are based on the assumption of the availability of quality information to model the physical elements in vehicular networks, without giving quantitative analysis on the quality of the logical views.

Several studies have evaluated the information quality in VCPS.
Liu et al. \cite{liu2014temporal} proposed a scheduling algorithm for temporal data dissemination in VCPS, which strikes a balance between real-time data dissemination and timely information sensing.
Dai et al. \cite{dai2019temporal} proposed an evolutionary multi-objective algorithm to enhance the information quality and improve the data delivery ratio.
Liu et al. \cite{liu2014realtime} proposed two online algorithms to schedule the temporal data dissemination under different consistency requirements by analyzing the dissemination characteristics.
Rager et al. \cite{rager2017scalability} developed a framework to enhance the information quality by modeling random data loads to capture the stochastic nature of real networks.
Yoon et al. \cite{yoon2021performance} presented a unified cooperative perception framework to obtain the accurate motion states of vehicles, considering communication losses in vehicular networks and the random vehicle motions.
These studies focused on information quality evaluation with respect to data timeliness, accuracy, or consistency in VCPS. Nevertheless, existing studies only considered the quality measurement at the homogeneous data item level, which may not sufficient when considering at the application level where the required logical views are constructed by fusing heterogeneous information.

Some studies have focused on vehicle sensing and information fusion by using DRL algorithms. 
Zhao et al. \cite{zhao2020social} designed a social-aware incentive mechanism based on proximal policy optimization (PPO) to derive the optimal long term sensing strategy.
Dong et al. \cite{dong2020spatio} presented a deep Q networks (DQN) based approach to fuse information obtained on the local downstream environment for reliable lane change  decisions.
Mika et al. \cite{mlika2022deep} proposed a deep deterministic policy gradient (DDPG) based solution to minimize the age of information by scheduling resource block and broadcast coverage.
These technologies are mainly proposed for vehicle sensing and information fusion using single agent DRL algorithms such as DQN, DDPG, and PPO. However, these algorithms cannot be directly applied for the cooperative sensing and heterogeneous information fusion in VCPS, and they are not suitable when considering multiple vehicles. A few studies applied the multi-agent DRL to allocate resources in vehicular networks.
Xu et al. \cite{xu2023joint} presented a multi-agent distributed distributional deep deterministic policy gradient (MAD4PG) to maximize the service ratio by scheduling the task offloading in vehicular edge computing. 
He et al. \cite{he2021efficient} proposed a multi-agent actor-critic (MAC) algorithm to allocate resources for vehicles with strict delay requirements and minimum bandwidth consumption.
Nevertheless, these solutions only considered one type of agents (i.e., vehicles or edge nodes) in vehicular networks.

\begin{figure*}
     \centering
     \subfloat[][Cooperative sensing and heterogeneous information fusion in VCPS]{\includegraphics[width=1.1\columnwidth]{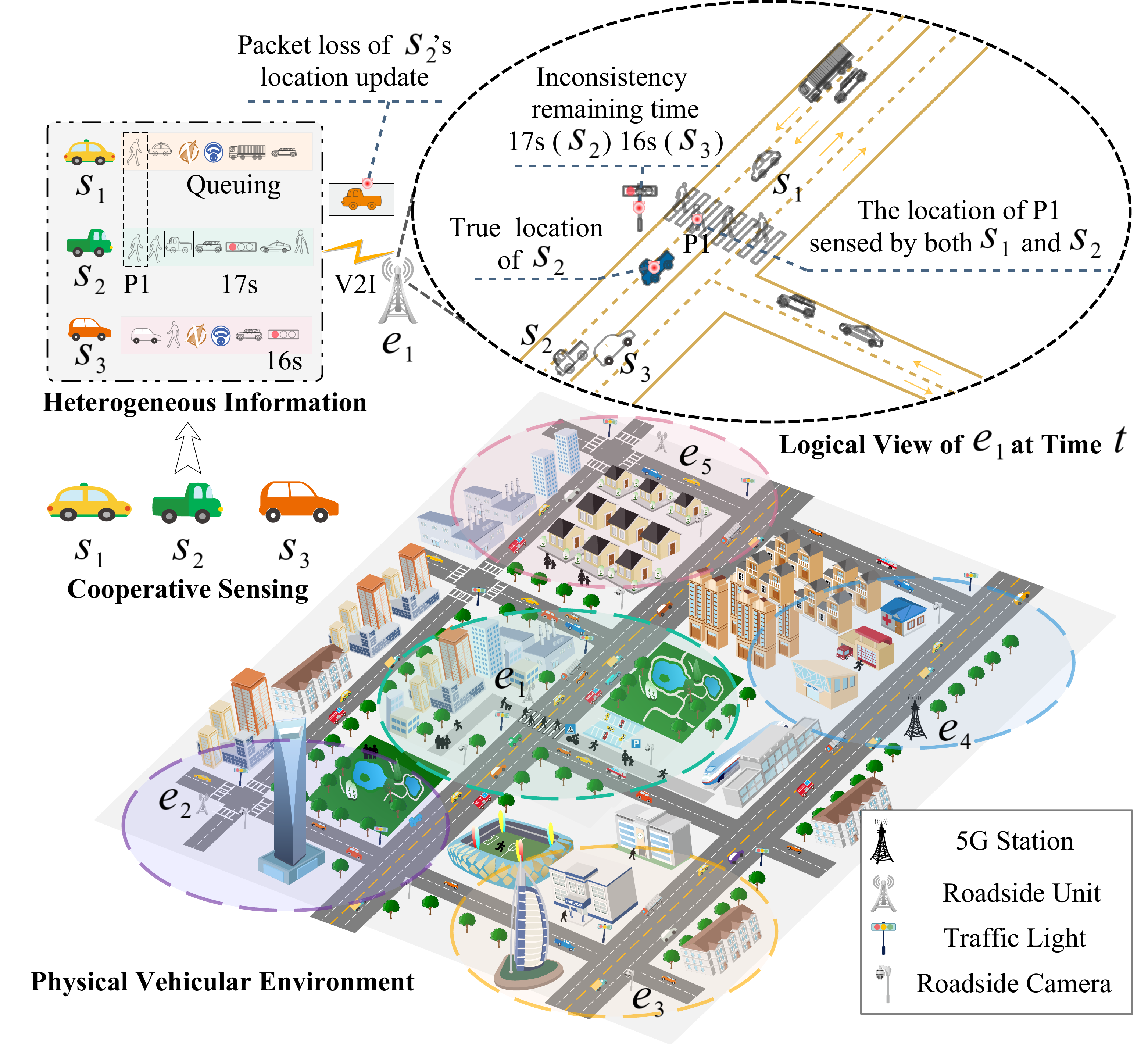}}
     \subfloat[][System workflow]{\includegraphics[width=0.8\columnwidth]{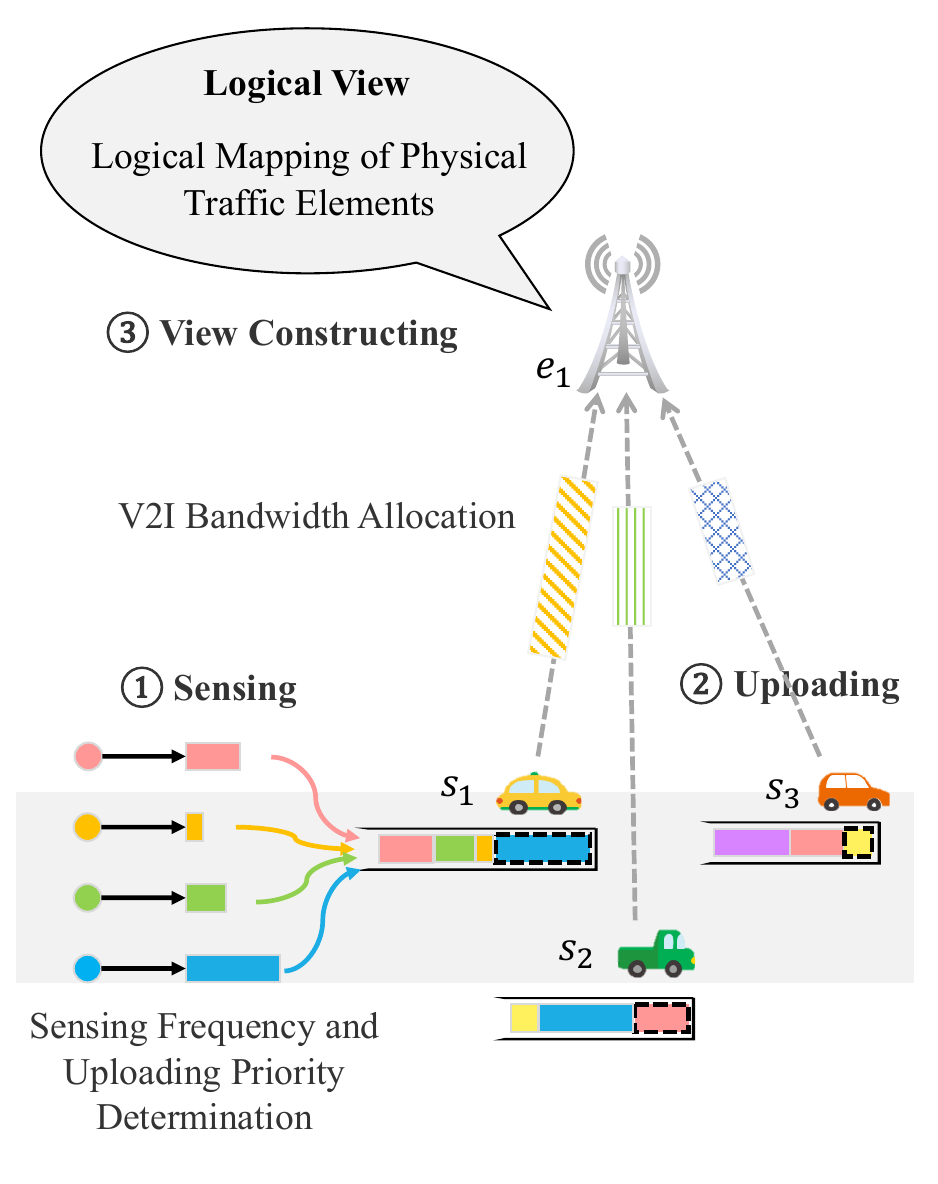}}
     \caption{System architecture}
     \label{fig_0_system_example}
\end{figure*}

\section{System Architecture}

In this section, we present a cooperative sensing and heterogeneous information fusion architecture in VCPS at the edge of vehicular networks.
As shown in Fig. \ref{fig_0_system_example}(a), the architecture can be abstracted into two layers, i.e., the physical vehicular environment and the logical views constructed by edge nodes.
In particular, edge nodes such as 5G stations and roadside units (e.g., $e_1$$\sim$$e_5$) are installed at the roadside.
Vehicles are able to communicate with edge nodes within their radio coverage via V2I communications and can sense heterogeneous information via equipped onboard sensors such as LiDAR, GPS, and cameras. 
Clearly, the physical information in vehicular networks are highly dynamic and temporal-spatial correlated. Meanwhile, the sensing vehicles are with heterogeneous capacities and limited resources, and the vehicular communications are intermittent and unreliable. Therefore, it is critical to have a tailored metric to quantitively evaluate the quality of the logical views constructed by the edge node, so as to measure the overall VCPS performance effectively.

As shown in Fig. \ref{fig_0_system_example}(b), the logical view construction at edge node $e_1$ consists of three steps. Step 1 (Sensing): Each vehicle may sense different information based on their locations and sensing capacities. The sensed information is queued at each vehicle for uploading to the edge node, and each vehicle will determine the sensing frequencies and uploading priorities of these information. Step 2 (Uploading): The edge node allocates V2I bandwidth (i.e., different ranges of non-overlapped spectrums) to vehicles with uploading tasks, so that these vehicles are able to upload their sensed information simultaneously without interference. Step 3 (View Constructing): The edge node constructs the logical view by mapping the received physical information to the corresponding logical elements based on the requirement of specific ITS applications.

The system characteristics are summarized as follows.
First, the heterogeneous information is sensed by vehicles at different sensing frequencies.
Therefore, the arrival moments of different information could be not the same. 
In addition, improving the sensing frequency may enhance information freshness, but also lengthen the queuing delay.
Second, it is essential to determine the uploading priorities of different information in vehicles by considering the different data sizes of information, connectivity of V2I communications, and view requirements comprehensively.
Third, due to the limited bandwidth resources of edge nodes and changeable vehicular channel conditions, the allocated V2I bandwidth may not be sufficient to support the data uploading timely.
It is meaningful to allocate larger bandwidth to vehicles that are prepared to upload fresher and more urgent information rather than in worse channel conditions (e.g., leaving the V2I coverage) to maximize bandwidth efficiency.
The channel conditions of different vehicles are considered by modeling the signal to noise ratio (SNR) between vehicles and the edge nodes, and the V2I transmission rate is determined by the distance between the two nodes and the allocated bandwidth.

Further, we give an example to better illustrate the idea.
As shown in Fig. \ref{fig_0_system_example}(a), a logical view is constructed in edge node $e_1$ at time $t$ to enable the speed advisory application at the intersection based on the information sensed and uploaded by vehicles $s_1$, $s_2$, and $s_3$.
In general, the goal of such an application is to advise optimal speed to the vehicles, which are approaching the intersection.
So, vehicles can pass smoothly and the overall traffic efficiency can be also maximized.
Suppose vehicles $s_2$ and $s_3$ can sense the traffic light information, but the values are not consistent at time $t$.
For example, $s_2$ observes 17s remaining of the red light, whereas $s_3$ observes 16s, resulting in the information inconsistency.
On the other hand, note that the status of the same physical element (e.g., the location of pedestrian P1) might be sensed by multiple vehicles simultaneously (e.g., $s_1$ and $s_2$). In such case, it only needs to be uploaded by one of the vehicles (e.g., vehicle $s_1$) at certain time to save the V2I bandwidth. As long as the physical elements are modeled at the edge node with the same quality level, it can be applied to different applications without the need of repeatedly uploading by different vehicles.
Moreover, the packet loss may cause a gap between the physical environment and the view.
For example, suppose the packet for $s_2$'s location update is lost, which results in the significant inconsistency between its true location and modeled location at time $t$.
As illustrated above, it is critical yet challenging to quantitatively measure the quality of views constructed at edge nodes, and design an effective scheduling mechanism for cooperative sensing and information fusion to maximize the overall quality of VCPS.

\section{Problem Formulation}

\subsection{Notations}
The set of discrete time slots of the system is denoted by $T=\left\{1,2,\cdots,t,\cdots,\left|T\right| \right\}$, where $\left|T\right|$ is the number of time slots.
The set of heterogeneous information is denoted by $D$.
Each information $d \in D$ is characterized by a two-tuple $d=\left(\operatorname{type}_d, \left|d\right| \right)$, where $\operatorname{type}_d$ is the type and $\left|d\right|$ is data size, measured by bit.
The set of vehicles is denoted by $S$.
Each vehicle $s \in S$ is characterized by a three-tuple $s=\left (l_s^t, D_s, \pi_s \right )$, where $l_s^t$ is the location of vehicle $s$ at time $t$; $D_s$ is the set of information that can be sensed by vehicle $s$, and $\pi_s$ is the transmission power of vehicle $s$.
The set of edge nodes is denoted by $E$.
Each edge node $e \in E$ is characterized by a three-tuple $e=\left (l_e, r_e, b_e \right)$, where $l_{e}$ is the location, $r_{e}$ is the communication range, and $b_{e}$ is the bandwidth capacity, measured by Hz.
The distance between vehicle $s$ and edge node $e$ at time $t$ is denoted by $\operatorname{dis}_{s,e}^t \triangleq \operatorname{distance} \left (l_s^t,l_e \right ), \forall s \in S, \forall e \in E, \forall t \in T$, where $\operatorname{distance}\left(\cdot,\cdot\right)$ is the Euclidean distance.

The set of information sensed by vehicle $s$ at time $t$ is denoted by $D_s^t \subseteq D_s$.
The information types are distinct for any information $d \in D_s^t$, i.e., $\operatorname{type}_{d^*} \neq \operatorname{type}_{d}, \forall d^* \in D_s^t \setminus \left\{ d\right \}, \forall d \in D_s^t$.
The sensing frequency of information $d$ in vehicle $s$ at time $t$ is denoted by $\lambda_{d,s}^t$.
Due to the limited sensing ability, we have $\lambda_{d,s}^{t} \in [\lambda_{d,s}^{\min} , \lambda_{d,s}^{\max} ], \ \forall d \in D_s^t, \forall s \in S, \forall t \in T$, where $\lambda_{d,s}^{\min}$ and $\lambda_{d,s}^{\max}$ are the minimum and maximum of sensing frequency for information with $\operatorname{type}_{d}$ in vehicle $s$, respectively.
The uploading priority of information $d$ in vehicle $s$ at time $t$ is denoted by $p_{d,s}^t$, and we have ${p}_{d^*, s}^t \neq {p}_{d, s}^t, \forall d^* \in D_s^t \setminus \left\{ d\right \}, \forall d \in D_s^t, \forall s \in S, \forall t \in T$.
The set of vehicles within the radio coverage of edge node $e$ at time $t$ is denoted by $S_e^t=\left \{s \vert \operatorname{dis}_{s,e}^t \leq r_e, \forall s \in S \right \}, S_e^t \subseteq S$.
The V2I bandwidth allocated by edge node $e$ for vehicle $s$ at time $t$ is denoted by $b_{s, e}^t$, and we have $b_{s, e}^t \in \left [0,b_e \right], \forall s \in S_e^{t}, \forall t \in T$.
The sum of V2I bandwidth allocated by edge node $e$ cannot exceed its capacity $b_e$, i.e., ${\sum_{\forall s \in S_e^{t}}b_{s, e}^t} \leq b_e, \forall t \in T$.
The primary notations are summarized in Table \ref{table_notations}.

\begin{table*}[ht]\small
\centering
\caption{Summary of primary notations}
\begin{tabular}[t]{lll}
\hline
\hline
Notations&Descriptions&Notes\\
\hline
$T$&Set of discrete time slots&$T=\left\{1,2,\cdots,t,\cdots,\left|T\right| \right\}$\\
$D$&Set of heterogeneous information& $d \in D$ and $d=\left(\operatorname{type}_d, \left|d\right| \right)$\\
$S$&Set of vehicles& $s \in S$ and $s=\left (l_s^t, D_s, \pi_s \right )$\\
$E$&Set of edge nodes& $e \in E$ and $e=\left (l_e, r_e, b_e \right)$\\
$\operatorname{type}_d$&Type of information $d$&\\
$l_{s}^{t}$&Location of vehicle $s$ at time $t$& \\
$D_s$&Set of information that can be sensed by vehicle $s$& \\
$\pi_{s}$&Transmission power of vehicle $s$&\\
$l_{e}$&Location of edge node $e$& \\
$r_{e}$&Communication range of edge node $e$& \\
$b_{e}$&Bandwidth capacity of edge node $e$& \\
$\operatorname{dis}_{s, e}^{t}$&Distance between vehicle $s$ and edge node $e$ at time $t$& $\operatorname{dis}_{s,e}^t \triangleq \operatorname{distance} \left (l_s^t,l_e \right )$\\
$S_{e}^{t}$&Set of vehicles within the radio coverage of edge node $e$& $S_e^t=\left \{s \vert \operatorname{dis}_{s,e}^t \leq r_e, \forall s \in S \right \}, S_e^t \subseteq S$\\
$D_s^t$&Set of information sensed by vehicle $s$ at time $t$&\\
$\lambda_{d, s}^{t}$&Sensing frequency of information $d$ in vehicle $s$ at time $t$& $\lambda_{d,s}^{t} \in \left [\lambda_{d,s}^{\min} , \lambda_{d,s}^{\max} \right ], \forall d \in D_s^t$\\
$p_{d, s}^{t}$&Uploading priority of information $d$ in vehicle $s$ at time $t$&${p}_{d^*, s}^t \neq {p}_{d, s}^t, \forall d^* \in D_s^t \setminus \left\{ d\right \}, \forall d \in D_s^t$\\
$b_{s, e}^{t}$&V2I bandwidth allocated by edge node $e$ for vehicle $s$ at time $t$&$b_{s, e}^t \in \left [0,b_e \right], \forall s \in S_e^{t}, \forall e \in E$\\
$\operatorname{a}_{d, s}^t$&Inter-arrival time of adjacent information with $\operatorname{type}_d$ in vehicle $s$ & \\
$D_{d, s}^t$&Set of elements with higher uploading priority than $d$ in vehicle $s$& $D_{d, s}^t = \left\{ d^* \mid p_{d^*,s}^{t} > p_{d,s}^{t} , \forall d^* \in D_s^t \right\}$\\
$\operatorname{q}_{d, s}^t$&Queuing time of information $d$ in vehicle $s$& \\
$\operatorname{g}_{d, s, e}^t$&Transmission time of information $d$ from vehicle $s$ to edge node $e$& \\
$\operatorname{c}_{d, s, e}^t$&Binary indicates whether $d$ is successfully transmitted from $s$ to $e$&$\operatorname{c}_{d, s, e}^t \in \{0, 1\}$ \\
$D_{s, e}^t$&Set of information transmitted by vehicle $s$ and received at $e$&$D_{s, e}^t = \left\{ d \mid \operatorname{c}_{d, s, e}^t = 1, \forall d \in D_s \right\}, D_{s, e}^t \subseteq D_s^t$\\ 
$O$&Set of physical traffic elements&\\
$V$&Set of views& $v = \left\{ o^{\prime} | y_{o,v} = 1, \forall o \in O \right\}$\\
$D_{v}$&Set of information required by view $v$&$D_{v} = \bigcup_{y_{o,v} = 1 \atop \forall o \in O} D_o, \forall v \in V$\\
$V_e^t$&Set of views required by edge node $e$ at time $t$& \\
$D_{v, e}$&Set of information received by edge node $e$ and required by view $v$&$D_{v, e}=\bigcup_{\forall s \in S}\left(D_v \cap D_{s, e}^t\right), \forall v \in V_e^t, \forall e \in E$\\
\hline
\hline
\end{tabular}
\label{table_notations}
\end{table*}

\subsection{Cooperative Sensing Model}
\begin{figure}
\centering
  \includegraphics[width=1\columnwidth]{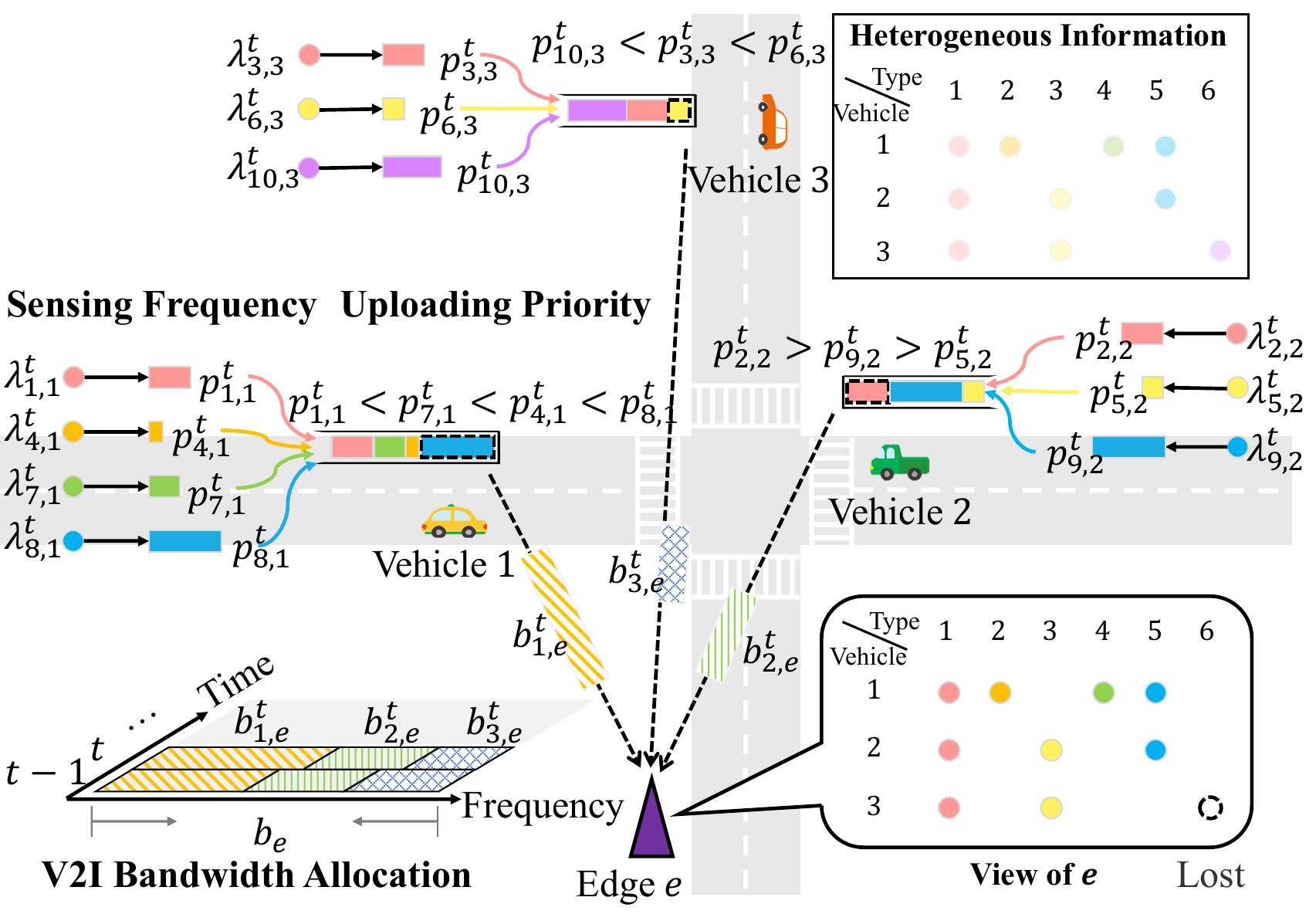}
  \caption{Cooperative sensing model}
  \label{fig_1_system_model}
\end{figure}

The cooperative sensing model is shown in Fig. \ref{fig_1_system_model}.
The inter-arrival time and queuing time of information sensed by vehicles are modeled by the multi-class M/G/1 priority queue \cite{qian2020minimizing}.
We assume that the distribution of information transmission time with the same $\operatorname{type}_d$ in vehicle $s$ stays stable within each time slot.
The transmission time $\operatorname{\hat{g}}_{d, s, e}^t$ of information with $\operatorname{type}_d$ follows a class of General distribution with mean $\alpha_{d, s}^t$, and finite second and third moments $\beta_{d, s}^t$, $\gamma_{d, s}^t$, and the distribution set is represented by 
\begin{equation}
\begin{aligned}
	\mathbb{P}=\left\{ \operatorname{\hat{g}}_{d, s, e}^t : \mathbb{E}\left[\operatorname{\hat{g}}_{d, s, e}^t\right] =\alpha_{d, s}^t, \right.\\ 
	 \mathbb{E} \left[\operatorname{\hat{g}}_{d, s, e}^t-\alpha_{d, s}^t\right]^{2}=\beta_{d, s}^t,  \\
	\left. \mathbb{E}\left[\operatorname{\hat{g}}_{d, s, e}^t-\alpha_{d, s}^t\right]^{3}=\gamma_{d, s}^t \right\}
\end{aligned}
\end{equation}
Therefore, the uploading workload $\rho_{s}^{t}$ is represented by 
\begin{equation}
    \rho_{s}^{t}=\sum_{\forall d \subseteq D_s^t} \lambda_{d,s}^{t}  \alpha_{d, s}^t
\end{equation}

To guarantee the existence of the queue steady-state, it requires $\rho_{s}^{t} < 1$.
The inter-arrival time $\operatorname{a}_{d, s}^t$ is the duration between the arrival of two adjacent information with $\operatorname{type}_d$ in vehicle $s$, which is computed by 
\begin{equation}
    \operatorname{a}_{d, s}^t=\frac{1}{\lambda_{d, s}^{t}}
\end{equation}
The set of elements with higher uploading priority than information $d$ in vehicle $s$ at time $t$ is denoted by
\begin{equation}
D_{d, s}^t=\left\{d^* \mid p_{d^*, s}^t>p_{d, s}^t, \forall d^* \in D_s^t\right\} 
\end{equation}
where $p_{d^*, s}$ is the uploading priority of information $d^* \in D_s^t$. Therefore, the uploading workload ahead of information $d$ (i.e., the amount of elements to be uploaded before $d$ by vehicle $s$ at time $t$) is represented by
\begin{equation}
\rho_{d, s}^t=\sum_{\forall d^* \in D_{d, s}^t} \lambda_{d^*, s}^t \alpha_{d^*, s}^t
\end{equation}
where $\lambda_{d^*, s}^t$ and $\alpha_{d^*, s}^t$ are the sensing frequency and the mean transmission time of information $d^*$ in vehicle $s$ at time $t$, respectively.

The queuing time of information with $\operatorname{type}_d$ in vehicle $s$ is denoted by $\operatorname{q}_{d, s}^t$. According to the Pollaczek$–$Khintchine formula \cite{takine2001queue}, the mean of queuing time $\operatorname{\bar{q}}_{d, s}^t$ is calculated by 
\begin{equation}
    \operatorname{\bar{q}}_{d, s}^t= \frac{1} {1 - \rho_{d,s}^{t}} 
        \left[ \alpha_{d, s}^t + \frac{ \lambda_{d,s}^{t} \beta_{d, s}^t + \sum\limits_{\forall d^* \in D_{d, s}^t} \lambda_{d^*,s}^t \beta_{d^*, s}^t }{2\left(1-\rho_{d,s}^{t} - \lambda_{d,s}^{t}  \alpha_{d, s}^t\right)}\right] 
        - \alpha_{d, s}^t
    \label{mean of queuing time}
\end{equation}
The variance of the queuing time of information with $\operatorname{type}_d$ in vehicle $s$ is obtained by Eq. \ref{variance of the queuing time}, where $\alpha_{d, s}^t$, $\beta_{d, s}^t$, and $\gamma_{d, s}^t$ are the mean, and finite second and third moments of transmission time of information $d$, respectively.
\begin{figure*}[hb]
 	\centering
 	\begin{equation}	    
 		\begin{aligned}
	{Var}(\operatorname{q}_{d, s}^t) &= \frac{\beta_{d, s}^t}{(1- \rho_{d,s}^{t})^2} + \frac{\alpha_{d, s}^t \sum\limits_{\forall d^* \in D_{d, s}^t} \lambda_{d^*,s}^t \beta_{d^*, s}^t}{(1- \rho_{d,s}^{t})^3} + \frac{\lambda_{d,s}^{t} \gamma_{d, s}^t + \sum\limits_{\forall d^* \in D_{d, s}^t} \lambda_{d^*,s}^t \gamma_{d^*, s}^t}{3(1- \rho_{d,s}^{t})^2(1-\rho_{d,s}^{t} - \lambda_{d,s}^{t}  \alpha_{d, s}^t)} \\
	&+ \frac{(\lambda_{d,s}^{t} \beta_{d, s}^t + \sum\limits_{\forall d^* \in D_{d, s}^t} \lambda_{d^*,s}^t \beta_{d^*, s}^t)^2}{4(1- \rho_{d,s}^{t})^2(1-\rho_{d,s}^{t} - \lambda_{d,s}^{t}  \alpha_{d, s}^t)^2}
	+ \frac{(\lambda_{d,s}^{t} \beta_{d, s}^t + \sum\limits_{\forall d^* \in D_{d, s}^t} \lambda_{d^*,s}^t \beta_{d^*, s}^t) \sum\limits_{\forall d^* \in D_{d, s}^t} \lambda_{d^*,s}^t \beta_{d^*, s}^t }{2(1- \rho_{d,s}^{t})^3(1-\rho_{d,s}^{t} - \lambda_{d,s}^{t}  \alpha_{d, s}^t)} - \beta_{d, s}^t
\end{aligned}
\label{variance of the queuing time}
 	\end{equation}
\end{figure*}
Based on Chebyshev's Inequality, we have the following inequality
\begin{equation}
	\operatorname{Pr}(|\operatorname{q}_{d, s}^t - \operatorname{\bar{q}}_{d, s}^t| > j \sqrt{{Var}(\operatorname{q}_{d, s}^t)}) \leq \frac{1}{j^2}, j \in \mathbb{R}^{+}
\end{equation}
Thus, the upper bound of queuing time under 99\% confidence level can be obtained by 
\begin{equation}
	\sup_{\operatorname{Pr}}{\operatorname{q}_{d, s}^t} \leq \operatorname{\bar{q}}_{d, s}^t + 10  \sqrt{{Var}(\operatorname{q}_{d, s}^t)}
\end{equation}

To better analyze the relationship between the mean queuing time and the uploading priorities of different elements in $D_s^t$, the Eq. 6 is rewritten as follows.
	\begin{equation}
\overline{\mathrm{q}}_{d, s}^t=\frac{\rho_{d, s}^t \alpha_{d, s}^t}{1-\rho_{d, s}^t}+\frac{\lambda_{d, s}^t \beta_{d, s}^t+\sum_{\forall d^* \in D_{d, s}^t}^t \lambda_{d^*, s}^t \beta_{d^*, s}^t}{2\left(1-\rho_{d, s}^t\right)\left(1-\rho_{d, s}^t-\lambda_{d, s}^t \alpha_{d, s}^t\right)}
\end{equation}
Suppose there are $n$ types of information, and the information ${d^1}$ has the highest uploading priority, i.e., $D_{d^1, s}^t = \emptyset$. Then, the mean queuing time of information ${d^1}$ can be computed by
\begin{equation}
\operatorname{\bar{q}}_{d^{1}, s}^t=\frac{\lambda_{d^1, s}^t \beta_{d^1, s}^t}{2}
\end{equation}
where $\lambda_{d^1, s}^t$ and $\beta_{d^1, s}^t$ are the sensing frequency and the second moment of transmission time of information $d^1$, respectively.
On the other hand, the information ${d^n}$ has the lowest uploading priority.
Since it requires that $\rho_s^t < 1$ to ensure the queue is stable and the queuing time is finite, we have
\begin{equation}
\rho_{d, s}^t=\sum_{\forall d^* \in D_{d, s}^t} \lambda_{d^*, s}^t \alpha_{d^*, s}^t<\sum_{\forall d \subseteq D_s^t} \lambda_{d, s}^t \alpha_{d, s}^t=\rho_s^t<1
\end{equation}
Similarly, $\rho_{d, s}^t+\lambda_{d, s}^t \alpha_{d, s}^t<1$. 
The mean queuing time of information ${d^n}$ can be obtained when $n$ tends to infinity due to $\lim _{n \rightarrow \infty}(1-\rho_{d^n, s}^t) \rightarrow 0$, similarly, $\lim _{n \rightarrow \infty}(1-\rho_{d^n, S}^t-\lambda_{d^n, s}^t \alpha_{d^n, s}^t) \rightarrow 0$.
\begin{equation}
\begin{aligned}
	\lim _{n \rightarrow \infty}\left(\mathrm{\bar{q}}_{d^n, s}^t\right)&=\frac{\lambda_{d^n, s}^t \beta_{d^n, s}^t+\sum_{\forall d^* \in D_{d^n, s}^t} \lambda_{d^*, s}^t \beta_{d^*, s}^t}{2\left(1-\rho_{d^n, s}^t\right)\left(1-\rho_{d^n, s}^t-\lambda_{d^n, s}^t \alpha_{d^n, s}^t\right)} \\
	&+ \frac{\rho_{d^n, s^t}^t \alpha_{d^n, s}^t}{1-\rho_{d^n, s}^t}\rightarrow \infty
\end{aligned}
\end{equation}
where $\lambda_{d^n, s}^t$, $\alpha_{d^n, s}^t$, and $\beta_{d^n, s}^t$ are the sensing frequency, the mean and second moment of transmission time of information $d^n$, respectively.

Then, we model the data uploading via V2I communications based on the Shannon theory.
The SNR of V2I communications between vehicle $s$ and edge node $e$ at time $t$ is denoted by $\operatorname{SNR}_{s,e}^{t}$, which is computed by \cite{sadek2009distributed}
\begin{equation}
    \label{equ_SNR}
    \operatorname{SNR}_{s,e}^{t}=\frac{1}{N_{0}}  \left|h_{s,e}\right|^{2} \zeta  {\operatorname{dis}_{s,e}^{t}}^{-\varphi} {\pi}_s
\end{equation}
where $N_{0}$ is the additive white Gaussian noise; $h_{s, e}$ is the channel fading gain; $\zeta$ is a constant that depends on the antennas design, and $\varphi$ is the path loss exponent.
Then, the V2I transmission rate between vehicle $s$ and edge node $e$ at time $t$, denoted by $\operatorname{z}_{s,e}^t$, is computed by 
\begin{equation}
    \operatorname{z}_{s,e}^t=b_{s, e}^{t} \log _{2}\left(1+\mathrm{SNR}_{s, e}^{t}\right)
\end{equation}
where $b_{s, e}^{t}$ is the bandwidth allocated to vehicle $s$ at time $t$.
Note that given the transmission power $\pi_s$ of vehicle $s$, the SNR of V2I communications between vehicle $s$ and edge node $e$ at time $t$ can be obtained by Eq. 14. Then, the transmission rate can be obtained by Eq. 15.

Assume vehicle $s$ is scheduled to upload $d$ at time $t$, and $d$ will be transmitted after a certain queuing time $\mathrm{\bar{q}}_{d, s}^t$. Then, we denote the moment when vehicle $s$ starts to transmit $d$ as $\mathrm{k}_{d, s}^t=t+\mathrm{q}_{d, s}^t$. The amount of data transmitted from time $\mathrm{k}_{d, s}^t$ to $\mathrm{k}_{d, s}^t + j$ can be obtained by $\int_{\mathrm{k}_{d, s}^t}^{\mathrm{k}_{d, s}^t+j} \mathrm{z}_{s, e}^t \mathrm{~d} t$ bits, where $j \in \mathbb{R}^{+}$ and $\mathrm{z}_{s, e}^t$ is the transmission rate at time $t$. If the amount of data could be transmitted during the entire transmission is larger than the size of $d$, the uploading would be complete. Therefore, the transmission time of information $d$ from vehicle $s$ to edge node $e$ is denoted by $\mathrm{g}_{d, s, e}^t$, which is computed by
	\begin{equation}
		\mathrm{g}_{d, s, e}^t=\inf _{j \in \mathbb{R}^{+}}\left\{\int_{\mathrm{k}_{d, s}^t}^{\mathrm{k}_{d, s}^t+j} \mathrm{z}_{s, e}^t \mathrm{~d} t \geq|d|\right\}
	\end{equation}

A successful transmission requires that the received SNR is above a certain threshold called SNR wall \cite{tandra2008snr} during the packet transmission, which is obtained by 
\begin{equation}
\mathrm{SNR}_{\text {wall }}=\frac{\sigma^{2}-1}{\sigma}
\end{equation}
where $\sigma=10^{\nu / 10}$, and $\nu$ is a parameter measured by dB that quantifies the size of the noise uncertainty, and $\left(\nu^2 - 1\right) {N_0} = {\pi_s} \nu $.
Thus, the successful transmission indicator, which indicates whether information $d$ is successfully transmitted from vehicle $s$ to edge node $e$, is represented by 
\begin{equation}
\label{Equ_pro_packet_loss}
\operatorname{c}_{d, s, e}^t=\left\{\begin{array}{l}
1, \forall {t^{*}} \in\left[\operatorname{k}_{d, s}^t, \operatorname{k}_{d, s}^t + \operatorname{g}_{d, s, e}^t\right], \operatorname{SNR}_{s,e}^{t^{*}}>\mathrm{SNR}_{\text {wall }}\\
0, \exists {t^{*}} \in\left[\operatorname{k}_{d, s}^t, \operatorname{k}_{d, s}^t + \operatorname{g}_{d, s, e}^t\right], \operatorname{SNR}_{s,e}^{t^{*}} \leq \mathrm{SNR}_{\text {wall }}
\end{array}\right.
\end{equation}
Therefore, the set of information transmitted by vehicle $s$ and received by edge node $e$ is denoted by $D_{s, e}^t = \{ d \mid \operatorname{c}_{d, s, e}^t = 1, \forall d \in D_s \}, D_{s, e}^t \subseteq D_s^t, \forall s \in S, \forall e \in E$.

\theoremstyle{definition}
\newtheorem{Definition}{Definition}

\subsection{Heterogeneous Information Fusion Model}
The set of physical elements in vehicular networks such as vehicles, pedestrians and roadside infrastructures is denoted by $O$. 
For each element $o \in O$, the edge node can construct a logical mapping $o^{\prime}$ based on the corresponding sensed information, and the set of information is denoted $D_o$. 
A view $v$ may contain multiple elements of the logical mapping based on specific ITS requirement, which is defined as follows.
\begin{Definition}[\textit{View $v$}]
View $v$ is a set of logical mapping of physical elements constructed at the edge node, which is constructed by fusing the sensed information of vehicles based on particular ITS application requirement and represented by
	\begin{equation}
		v = \left\{ o^{\prime} | y_{o,v} = 1, \forall o \in O \right\}
	\end{equation}
where $y_{o, v} \in \{0, 1\}$ indicates whether the logical mapping of the physical element $o$ is included in view $v$.
\end{Definition}
The set of information required by view $v$ is denoted by $D_{v}$, which is the mapping of physical traffic elements required by particular ITS application, and it is represented by
\begin{equation}
	D_{v} = \bigcup_{y_{o,v} = 1 \atop \forall o \in O} D_o
\end{equation}
The number of required elements in view $v$ is denoted by $|D_{v}|$.
Denote the set of views in the system as $V$, and denote the set of views required by edge node $e$ at time $t$ by $V_e^t \subseteq V$.
Thus, the set of information received by edge node $e$ and required by view $v$ is represented by 
\begin{equation}
    D_{v, e}=\bigcup_{\forall s \in S}\left(D_v \cap D_{s, e}^t\right), \forall v \in V_e^t, \forall e \in E
\end{equation}
and $| D_{v, e} |$ is the number of information that received by edge node $e$ and required by view $v$.
Then, we define the three characteristics of heterogeneous information fusion, including the timeliness, completeness, and consistency of the view as follows.

First, the heterogeneous information is time-varying, and information freshness is essential for modeling the quality of views.
Thus, we define the timeliness of information $d$ in vehicle $s$ as follows.
\begin{Definition}[\textit{Timeliness of information $d$}]
	The timeliness $\xi_{d,s} \in (0, +\infty)$ of information $d$ in vehicle $s$ is defined as the sum of the intel-arrival time, queuing time, and transmission time of the information $d$.
	\begin{equation}
    	\xi_{d,s} = \operatorname{a}_{d,s}^t + \operatorname{q}_{d, s}^t + \operatorname{g}_{d, s, e}^t, \forall d \in D_s^t,\forall s \in S
    	\label{equ_timeliness_of_information}
	\end{equation}
\end{Definition}
\noindent where $\operatorname{a}_{d,s}^t$, $\operatorname{q}_{d, s}^t$ and $\operatorname{g}_{d, s, e}^t$ are the intel-arrival time, queuing time and transmission time of information $d$, respectively.
Therefore, we define the timeliness of a view as follows.
\begin{Definition}[\textit{Timeliness of view $v$}]
	The timeliness $\Xi_{v} \in (0,+\infty)$ of view $v$ is defined as the sum of the information timeliness.
	\begin{equation}
    	\Xi_{v} = \sum_{\forall s\in S} \sum_{\forall d \in D_{v, e} \cap D_s^t } \xi_{d,s}, \forall v \in V_e^t, \forall e \in E
    	\label{equ_timeliness}
	\end{equation}
\end{Definition}

Second, vehicular networks have several intrinsic characteristics including the high mobility of vehicles, restricted network resources, and unreliable wireless communications. 
The view may be incomplete due to the disconnection between the vehicle and the edge node, or the loss of the packet. 
Therefore, we define the completeness of a view as follows.
\begin{Definition}[\textit{Completeness of view $v$}]
	The completeness $\Phi_{v} \in [0,1]$ of view $v$ is defined as the ratio of the number of information actually received by edge node $e$ to the number of total required information.
	\begin{equation}
	\Phi_{v}= {| D_{v, e} |} \big/ {|D_{v} |}, \forall v \in V_e^t, \forall e \in E
	\end{equation}
\end{Definition}
\noindent where $| D_{v, e} |$ is the number of information received by edge node $e$ and required by view $v$, and $|D_{v}|$ is the number of information required by view $v$.

Third, since different types of information have their own sensing frequencies and uploading priorities, it is important to keep the versions of different types of information as close as possible when constructing a view. 
Therefore, we define the consistency of a view as follows. 
\begin{Definition}[\textit{Consistency of view $v$}]
	 The consistency $\Psi_{v} \in (0,+\infty)$ of view $v$ is defined as the quadratic sum of the difference between the receiving time of information $d$ and average receiving time of  information required by the view.
\begin{equation}
\Psi_{v}=\sum_{\forall s\in S} \sum_{\forall d \in D_{v, e} \cap D_s^t} \left|\operatorname{q}_{d, s}^t + \operatorname{g}_{d, s, e}^t - \psi_{v} \right|^{2}, \forall v \in V_e^t, \forall e \in E
\end{equation}
\end{Definition}
\noindent where $\psi_{v}$ is average receiving time of information required by view $v$, which is computed by 
\begin{equation}
	\psi_{v} = \frac{1}{|D_{v, e}|} {\sum_{\forall s \in S}\sum_{\forall d \in D_{v, e} \cap D_s^t} \left( \operatorname{q}_{d, s}^t + \operatorname{g}_{d, s, e}^t\right) }, \forall v \in V_e^t, \forall e \in E
\end{equation}

Finally, we give the formal definition of the age of view, which synthesizes the timeliness, completeness, and consistency of the view.
\begin{Definition}[\textit{Age of View, AoV}]
	The age of view $\operatorname{AoV}_{v} \in (0, 1)$ is defined as the weighted average of normalized timeliness, completeness and consistency of the view $v$.
	\begin{equation}
	    \operatorname{AoV}_{v} = w_1  \hat{\Xi}_{v} + w_2  \hat{\Phi}_{v}+  w_3 \hat{\Psi}_{v}, \forall v \in V_e^t, \forall e \in E
\end{equation}
\end{Definition}
\noindent where $\hat{\Xi}_{v} \in (0, 1)$, $\hat{\Phi}_{v} \in (0, 1)$ and $\hat{\Psi}_{v} \in (0, 1)$ denote the normalized timeliness, normalized completeness, and normalized consistency of view $v$, respectively.
Note that since the dimensions of the timeliness, completeness, and consistency of view are different, in order to form a uniform representation of AoV, they are normalized to $(0,1)$ based on the min-max scaling as follows.
\begin{equation}
    \begin{cases}
		\hat{\Xi}_{v} = {\Xi}_{v} \big/ \left( \delta_\xi | D_{v, e} |   |T| \right)  \\ 
		\hat{\Phi}_{v} = 1 - {\Phi}_{v}  \\
		\hat{\Psi}_{v} = {\Psi}_{v} \big/ \left( \delta_\psi  \max\limits_{\substack{\forall d \in D_v \cap D_s^t \\ \forall s \in S}}{\left\{ \left|\operatorname{q}_{d, s}^t + \operatorname{g}_{d, s, e}^t - \psi_{v} \right|^{2}\right\}}   \right)
	\end{cases}
\end{equation}
\noindent where $\delta_{\xi} \in(0,1)$ and $\delta_\psi \in(0,1)$ are the data scaling factors of the timeliness and consistency, respectively. They are utilized to avoid concentrating most of the values in a small range by scaling down the theoretical maximum of the timeliness and consistency in the min-max scaling.

The weighting factors of $\hat{\Xi}_{v}$, $\hat{\Phi}_{v}$, and $\hat{\Psi}_{v}$ are denoted by $w_1$, $w_2$, and $w_3$, respectively, and $w_1+w_2+w_3=1$.
These three weighting factors can be tuned accordingly based on the different requirements of ITS applications.
For example, for the speed advisory application at the road intersection, the vehicles are expected to receive instructions of real-time velocity from the edge node so as to pass the intersection safely and smoothly. In such a case, the timeliness factor (e.g., real-time traffic light status is more important to be modeled in the view compared with the completeness factor (e.g., the pedestrians to be modeled in the view).
Note that the lower value of $\operatorname{AoV}_{v}$ indicates higher quality of the constructed view.

\subsection{Quality of VCPS}
Given the above metric AoV, which evaluates the quality of views individually, we further define the quality of VCPS at the system level as follows.
\begin{Definition}[\textit{VCPS Quality}]
	 The quality of VCPS $\Upsilon \in (0,1)$ is defined as the average of the complement of AoV for each view $v$ in edge nodes during the scheduling period $T$.
\begin{equation}
\Upsilon=\frac{\sum_{\forall t \in T} \sum_{\forall e \in E} \sum_{\forall v \in V_e^t} \left(1 - \operatorname{AoV}_{v}\right)}{\sum_{\forall t \in T} \sum_{\forall e \in E} |V_e^t| }
\end{equation}
\end{Definition}

Given a solution $(\bf\Lambda, \mathbf{P}, \mathbf{B} )$, where $\bf\Lambda$ denotes the determined sensing frequencies, $\mathbf{P}$ denotes the determined uploading priorities, and $\mathbf{B}$ denotes the determined V2I bandwidth allocation, which are represented by 
\begin{equation}
    \begin{cases}
\bf\Lambda &= \left \{ \lambda_{d,s}^{t} \vert \forall d \in D_s^t  , \forall s \in S, \forall t \in T \right \} \\ 
\mathbf{P} &= \left \{ p_{d,s}^{t} \vert \forall d \in D_s^t  , \forall s \in S, \forall t \in T\right \}  \\
\mathbf{B} &= \left \{ b_{s, e}^t \vert \forall s \in S_e^t, \forall e \in E, \forall t \in T\right \}
\end{cases}
\end{equation}
\noindent the problem is to maximize the quality of VCPS, which is expressed as follows:
\begin{equation}
	\begin{aligned}
		&\max_{\bf\Lambda, \mathbf{P}, \mathbf{B}} \Upsilon \\
		\text { s.t. }
        \operatorname{C1}: & \lambda_{d,s}^{t} \in \left [\lambda_{d,s}^{\min} , \lambda_{d,s}^{\max} \right ], \forall d \in D_s^t , \forall s \in S, \forall t \in T\\
        \operatorname{C2}: &{p}_{d^*, s}^t \neq {p}_{d, s}^t, \forall d^* \in D_s^t \setminus \left\{ d\right \}, \forall d \in D_s^t, \forall s \in S, \forall t \in T\\
        \operatorname{C3}: & b_{s, e}^t \in \left[ 0 , b_e \right ], \forall s \in S_e^t, \forall e \in E, \forall t \in T\\
        \operatorname{C4}: & \sum_{\forall d \subseteq D_s^t} \lambda_{d,s}^{t} \cdot \alpha_{d, s}^t < 1,\ \forall s \in S, \forall t \in T \\
        \operatorname{C5}: & {\sum_{\forall s \in S_e^{t}}b_{s, e}^t} \leq b_e, \forall e \in E, \forall t \in T
	\end{aligned}
\end{equation}

Constraint $\operatorname{C1}$ requires that the sensing frequencies of information $d$ in vehicle $s$ at time $t$ should meet the requirement of its sensing ability.
$\operatorname{C2}$ guarantees the uploading priority of information $d$ in vehicle $s$ at time $t$.
$\operatorname{C3}$ states that the V2I bandwidth allocated by the edge node $e$ for vehicle $s$ at time $t$ cannot exceed its bandwidth capacity $b_e$.
$\operatorname{C4}$ guarantees the queue steady-state during the scheduling period $T$. 
$\operatorname{C5}$ requires that the sum of V2I bandwidth allocated by the edge node $e$ cannot exceed its capacity $b_e$.

\section{Proposed Solution}

\subsection{Solution Model}
\begin{figure*}
\centering
  \includegraphics[width=1.4\columnwidth]{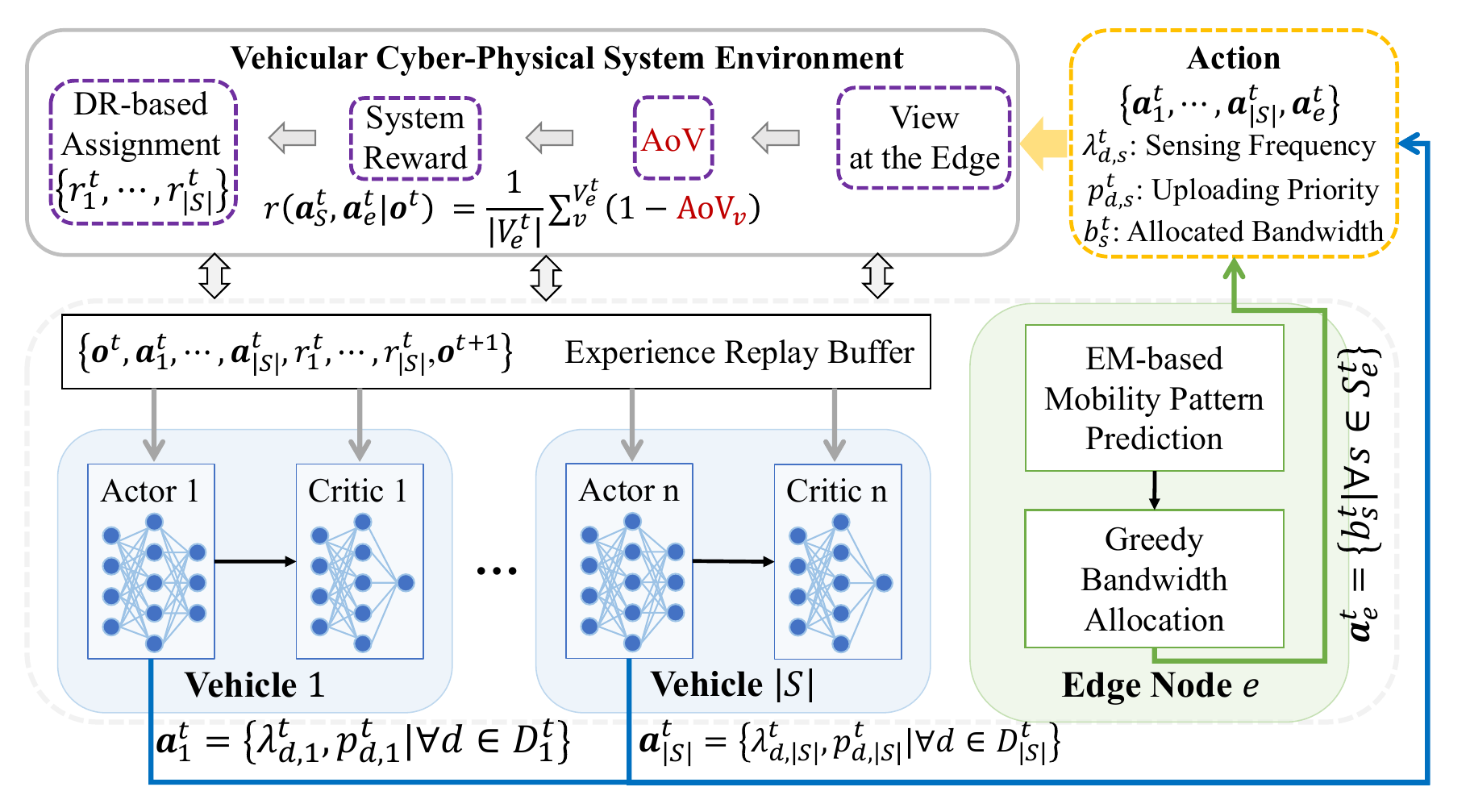}
  \caption{Model of the proposed solution}
  \label{fig_3_HRL}
\end{figure*}

As shown in Fig. \ref{fig_3_HRL}, the solution model consists of $|S|$ vehicles, edge node $e$, the VCPS environment, and an experience replay buffer.
First, vehicle $s$ decides its action $\boldsymbol{a}_{s}^{t}$ on determining the sensing frequencies and uploading priorities.
In particular, the actor network of vehicle $s$ is utilized to generate its action with the input of the local observation $\boldsymbol{o}_{s}^{t}$ of the system state.
The critic network of vehicle $s$ is utilized to evaluate the action generated by the corresponding actor network.
Second, edge node $e$ decides its action $\boldsymbol{a}_{e}^{t}$ on allocating V2I bandwidth for vehicles within its radio coverage  based on predicted vehicle trajectories and view requirements.
Third, the environment obtains the system reward according to the actions $\{ \boldsymbol{a}_{1}^{t}, \cdots, \boldsymbol{a}_{|S|}^{t}, \boldsymbol{a}_{e}^{t}\}$.
The system reward is the achieved VCPS quality in edge node $e$ at time $t$.
A DR-based credit assignment is adopted to divide the system reward into the difference reward $\{r_1^t, \cdots, r_{|S|}^t\}$, where $r_s^t$ is utilized to evaluate the contribution of vehicle $s$ on view construction.
Fourth, the related interaction experiences including the system state, vehicle actions, difference reward and next system state are stored in the experience replay buffer, and they are utilized to train the actor and critic networks in vehicles.

The primary components of the solution model are designed as follows.

\textit{1) System State:} The edge node broadcasts its view requirements and cached information periodically. The local observation of the system state in vehicle $s$ at time $t$ is denoted by 
	\begin{equation}
		\boldsymbol{o}_{s}^{t}=\left\{D_{s}^{t}, D_{e}^{t}, V_e^t\right\}
	\end{equation} 
	\noindent where $D_{s}^{t}$ represents the set of information sensed by vehicle $s$ at time $t$,
	$D_{e}^{t}$ represents the set of cached information in edge node $e$ at time $t$,
	and $V_e^t$ represents the set of views required by edge node $e$ at time $t$.
	The system state at time $t$ is denoted:
	\begin{equation}
		\boldsymbol{o}^{t}=\left\{D_{1}^{t}, \cdots, D_{s}^{t}, \cdots, D_{|S|}^{t}, D_{e}^{t}, V_{e}^{t}\right\}
	\end{equation}

\textit{2) Action Space:} The action space of vehicle $s$ consists of the sensing frequencies and uploading priorities of the sensed information at time $t$, which is denoted by 
	\begin{equation}
		\boldsymbol{a}_{s}^{t}=\{ \lambda_{d, s}^{t}, p_{d, s}^{t} \mid \forall d \in D_{s}^t\}
	\end{equation}
	\noindent where $\lambda_{d, s}^{t}$ and $p_{d, s}^{t}$ are the sensing frequency and uploading priority of information $d$ in vehicle $s$ at time $t$, respectively.
	The set of vehicle actions is denoted by $\boldsymbol{a}_{S}^{t} = \left\{\boldsymbol{a}_{s}^{t}\mid \forall s \in S\right\}$.
	The action of edge node is the V2I bandwidth allocation for vehicles, which is denoted by 
	\begin{equation}
		\boldsymbol{a}_{e}^{t}=\{b_{s, e}^{t} \mid \forall s \in S_{e}^{t}\}
	\end{equation}
	where $b_{s, e}^t$ is the V2I bandwidth allocated by edge node $e$ for vehicle $s$ at time $t$.
	
\textit{3) System Reward:} The system reward via vehicle actions $\boldsymbol{a}_{S}^{t}$ and edge action $\boldsymbol{a}_{e}^{t}$ in the system state $\boldsymbol{o}^{t}$ is defined as the achieved VCPS quality in edge node $e$ at time $t$, which is computed by 
	\begin{equation}
		r\left(\boldsymbol{a}_{S}^{t},\boldsymbol{a}_{e}^{t} \mid \boldsymbol{o}^{t}\right)=\frac{1}{\left|V_e^t\right|} \sum_{\forall v \in V_e^t}\left(1 -\operatorname{AoV}_{v} \right)
	\end{equation}

The system reward reflects the entire system performance, which is contributed by both vehicles and the edge node. 
It is expected to assign the system reward into individual rewards for vehicles to evaluate their contributions.
The DR-based credit assignment scheme use the difference between the system reward and the reward achieved without the agent action to represent the individual reward of the agent, so that it can further improve the performance of the proposed solution by accurately evaluating the action of each agent separately.
Accordingly, the difference reward of vehicle $s$ is represented by \cite{foerster2018counterfactual}
\begin{equation}
\label{equa_dr_sensor_node}
r_{s}^{t}=r\left(\boldsymbol{a}_{S}^{t},\boldsymbol{a}_{e}^{t} \mid \boldsymbol{o}^{t}\right)-r\left(\boldsymbol{a}_{S-s}^{t},\boldsymbol{a}_{e}^{t} \mid \boldsymbol{o}^{t}\right)
\end{equation}
\noindent where $r\left(\boldsymbol{a}_{S-s}^{t},\boldsymbol{a}_{e}^{t} \mid \boldsymbol{o}^{t}\right)$ is the system reward achieved without the contribution of vehicle $s$, and it can be obtained by setting null action set for vehicle $s$.
The set of the difference reward for vehicles is denoted by $\boldsymbol{r}_{S}^{t}=\{ r_{s}^{t} \mid \forall s \in S\}$.

\subsection{Workflow}
The proposed solution includes three parts, namely, the initialization, replay experiences storing and training. 

1) \textbf{Initialization}: Each vehicle consists of four neural networks, i.e., a local actor, a target actor, a local critic, and a target critic network.
The parameters of the local actor and local critic networks of vehicle $s$ are denoted by $\theta_{s}^{\mu}$ and $\theta_{s}^{Q}$, respectively.
The parameters of target actor and target critic networks are denoted by $\theta_{s}^{\mu^{\prime}}$ and $\theta_{s}^{Q^{\prime}}$, respectively.
The parameters of local actor and local critic networks of vehicles are randomly initialized.
The parameters of target actor and target critic networks are initialized as the same as the corresponding local network,
\begin{equation}
	\begin{aligned}
		\theta_{s}^{\mu^{\prime}} \leftarrow \theta_{s}^{\mu}, \forall s \in S\\
		\theta_{s}^{Q^{\prime}} \leftarrow \theta_{s}^{Q}, \forall s \in S
	\end{aligned}
\end{equation}
An experiment replay buffer $\mathcal{B}$ with a maximum size $|\mathcal{B}|$ is initialized to store replay experiences for vehicles.
The procedure is shown in lines 1$\sim$2 in Algorithm 1.

2) \textbf{Replay experiences storing}: 
At the beginning of each iteration, a random process $\mathcal{N}$ is initialized for exploration.
The action of vehicle $s$ at time $t$ is obtained by the local actor network based on its local observation of the system state.
\begin{equation}
	\boldsymbol{a}_{s}^{t}=\boldsymbol{\mu}_{\boldsymbol{s}}\left(\boldsymbol{o}_{s}^{t} \mid \theta_{s}^{\mu}\right)+\mathcal{N}_{t}
\end{equation}
\noindent where $\mathcal{N}_{t}$ is an exploration noise to increase the diversity of vehicle actions.

Edge node $e$ allocates the V2I bandwidth via the VBA scheme based on predicted vehicle trajectories and view requirements.
First, the mobility patterns of vehicles are predicted by edge node $e$ using the expectation-maximization (EM) method \cite{hofmann2001unsupervised} based on historical distances between vehicles and the edge node.
Then, the trajectory of vehicle $s$ at future $h$ time slots is predicted based on the EM-based mobility pattens prediction, which is denoted by $\operatorname{Traj}_{s}^{t} = \{ \hat{l}_{s}^{t+1}, \hat{l}_{s}^{t+2}, \dots, \hat{l}_{s}^{t+h}\}$, where $\hat{l}_{s}^{t+1}$ is the predicted location of vehicle $s$ at time $t+1$.
Thus, he average distance of vehicle between edge node is computed by 
\begin{equation}
	\operatorname{\bar{dis}}_{s, e}^{t} = \frac{1}{h} {\sum_{\forall i \in [1, h]} \widehat{\operatorname{dis}}_{s, e}^{t+i}}
\end{equation}
where $\widehat{\operatorname{dis}}_{s, e}^{t+i}$ is the distance between the predicted location of vehicle $s$ and the edge node, and $\widehat{\operatorname{dis}}_{s, e}^{t+i}=\operatorname{distance}(\hat{l}_{s}^{t+i}, l_{e})$.

\begin{algorithm}[t]
		\caption{Proposed Solution}
		Initialize the parameters;\\
		Initialize the experience replay buffer $\mathcal{B}$;\\
        \For{iteration $= 1$ to max-iteration-number}{
            Initialize a random process $\mathcal{N}$ for exploration; \\
            Receive the initial system state $\boldsymbol{o}_{1}$;\\
            \For{$t = 1$ to $|T|$}{
            	\For{vehicle $s=1$ to $|S|$ }{
            			Receive a local observation $\boldsymbol{o}_{s}^{t}$; \\
                    	Select a action $\boldsymbol{a}_{s}^{t}=\boldsymbol{\mu}_{\boldsymbol{s}}\left(\boldsymbol{o}_{s}^{t} \mid \theta_{s}^{\mu}\right)+\mathcal{N}_{t}$; \\
            		Obtain the required information $D_{s,\operatorname{R}}^{t}$; \\
        		Predict the mobility patten via EM-based method with historical relative distances; \\
        		Predict the future locations $\operatorname{Traj}_{s}^{t}$; \\
        		Compute the average distance $\operatorname{\bar{dis}}_{s, e}^{t}$;
            	}
        	\For{vehicle $s=1$ to $|S|$ }{
        		Allocate the bandwidth $b_{s, e}^{t}$ to $s$ via the VBA scheme;\\}
            	Receive the system reward $r\left(\boldsymbol{a}_{S}^{t},\boldsymbol{a}_{e}^{t} \mid \boldsymbol{o}^{t}\right)$ and the next system state $\boldsymbol{o}^{t+1}$;\\
            	Divide the system reward into the difference reward $\boldsymbol{r}_{S}^{t}$ for vehicles;\\
            	Store $\left(\boldsymbol{o}^{t}, \boldsymbol{a}_{S}^{t}, \boldsymbol{r}_{S}^{t}, \boldsymbol{o}^{t+1}\right)$ in replay buffer $\mathcal{B}$;
            }
            \For{vehicle $s=1$ to $|S|$ }{
            		Sample $M$ transitions from $\mathcal{B}$ randomly;\\
            		Update the local critic and local actor networks;\\
            	}
            	Update the target networks of vehicles;
       	}
\end{algorithm}

Then, the set of information sensed by vehicle $s$ and required by view $v$ at time $t$ is represented by 
\begin{equation}
	D_{s, v}^{t} = \left\{ d \mid  d \in D_{s}^t \cap  D_v \right\}
\end{equation}
Thus, the set of information sensed by vehicle $s$ and required by views in edge node $e$ at time $t$ is denoted by 
\begin{equation}
	D_{s, {V_e^t}}^{t} = \{ d \mid  d \in \bigcup_{\forall v \in V_e^t} D_{s, v}^{t}\}
\end{equation}
\noindent and its size is denoted by $|D_{s, {V_e^t}}^{t}|$, which is computed by 
\begin{equation}
	|D_{s, {V_e^t}}^{t}| = \sum_{\forall d \in D_{s, {V_e^t}}^{t}}|d|
\end{equation}
Finally, the V2I bandwidth allocated by edge node $e$ for vehicle $s$ is calculated by 
\begin{equation}
	b_{s, e}^{t} =\frac{b_{e}} {\omega+\operatorname{rank}_{s}}
\end{equation}
\noindent where $\omega$ is a constant, and $\operatorname{rank}_{s}$ is the sort ranking of vehicle $s$ by the sequence of $| D_{s, {V_e^t}}^{t}|$ in the descending order and $\operatorname{\bar{dis}}_{s, e}^{t}$ in the ascending order.

After determining the actions of vehicles and the edge node, the system reward $r\left(\boldsymbol{a}_{S}^{t},\boldsymbol{a}_{e}^{t} \mid \boldsymbol{o}^{t}\right)$ is obtained as the achieved VCPS quality, which is further divided into the difference reward $\boldsymbol{r}_{S}^{t}$ via the DR-based credit assignment scheme.
Finally, the interaction experiences including the system state $\boldsymbol{o}^{t}$, vehicle actions $\boldsymbol{a}_{S}^{t}$, difference reward $\boldsymbol{r}_{S}^{t}$, and next system state $\boldsymbol{o}^{t+1}$, are stored in the experience replay buffer $\mathcal{B}$.
The procedure is shown in lines 4$\sim$18 in Algorithm 1.

3) \textbf{Training}: A minibatch of $M$ transitions is sampled from experience replay buffer $\mathcal{B}$ to train the actor and critic networks in vehicles. 
The transition of the $M$ minibatch is denoted by $(\boldsymbol{o}_{s}^{m}, \boldsymbol{a}_{S}^{m}, \boldsymbol{r}_{S}^{m}, \boldsymbol{o}_{s}^{m+1})$.
The loss function of the local critic network of vehicle $s$ is computed by 
\begin{equation}
\label{equ_loss_sensor}
	\mathcal{L}\left(\theta_{s}^{Q}\right)=\frac{1}{M} \Sigma_{m}\left(\eta_{m}-Q_{s}\left(\boldsymbol{o}_{s}^{m}, \boldsymbol{a}_{S}^{m} \mid \theta_{s}^{Q}\right)\right)^{2}
\end{equation}
\noindent where $\eta_{m}$ is the target value generated by the target critic network $\eta_{m}=r_{s}^{m}+\tau Q_{s}^{\prime}(\boldsymbol{o}_{s}^{m+1}, \boldsymbol{a}_{S}^{m+1} \mid \theta_{s}^{Q^{\prime}})$, and $\tau$ is the discount rate. 
The action of vehicle $s$ at time $m+1$ is obtained by the target actor network based on the local observation of next system state, i.e., $\boldsymbol{a}_{s}^{m+1}=\mu_{s}^{\prime}(\boldsymbol{o}_{s}^{m+1} \mid \theta_{s}^{\mu^{\prime}})$.
The parameters of the local actor network of vehicle $s$ are updated via policy gradient.
\begin{equation}
\label{equ_gradient_sensor}
	\text{\footnotesize$\nabla_{\theta_{s}^{\mu}} \mathcal{J} \approx \frac{1}{M} \sum_{m} \nabla_{\boldsymbol{a}_{s}^{m}} Q_{s}\left(\boldsymbol{o}_{s}^{m}, \boldsymbol{a}_{S}^{m} \mid \theta_{s}^{Q}\right) \nabla_{\theta_{s}^{\mu}} \mu_{s}\left(\boldsymbol{o}_{s}^{m+1} \mid \theta_{s}^{\mu}\right)$}
\end{equation}
Finally, vehicles update the parameters of target networks,
\begin{equation}
\label{equ_update_target}
	\begin{aligned}
			\theta_{s}^{\mu^{\prime}} &\leftarrow n_{s} \theta_{s}^{\mu}+(1-n_{s})  \theta_{s}^{\mu^{\prime}}, \forall s \in S\\
			\theta_{s}^{Q^{\prime}} &\leftarrow n_{s} \theta_{i}^{Q}+(1-n_{s})  \theta_{s}^{Q^{\prime}}, \forall s \in S
	\end{aligned}
\end{equation}
\noindent with $n_{s} \ll 1, \forall s \in S $.
The procedure is shown in lines 19$\sim$22 in Algorithm 1.

\section{Performance Evaluation}

\subsection{Settings}
\begin{figure*}
\centering
  \includegraphics[height=2.12in]{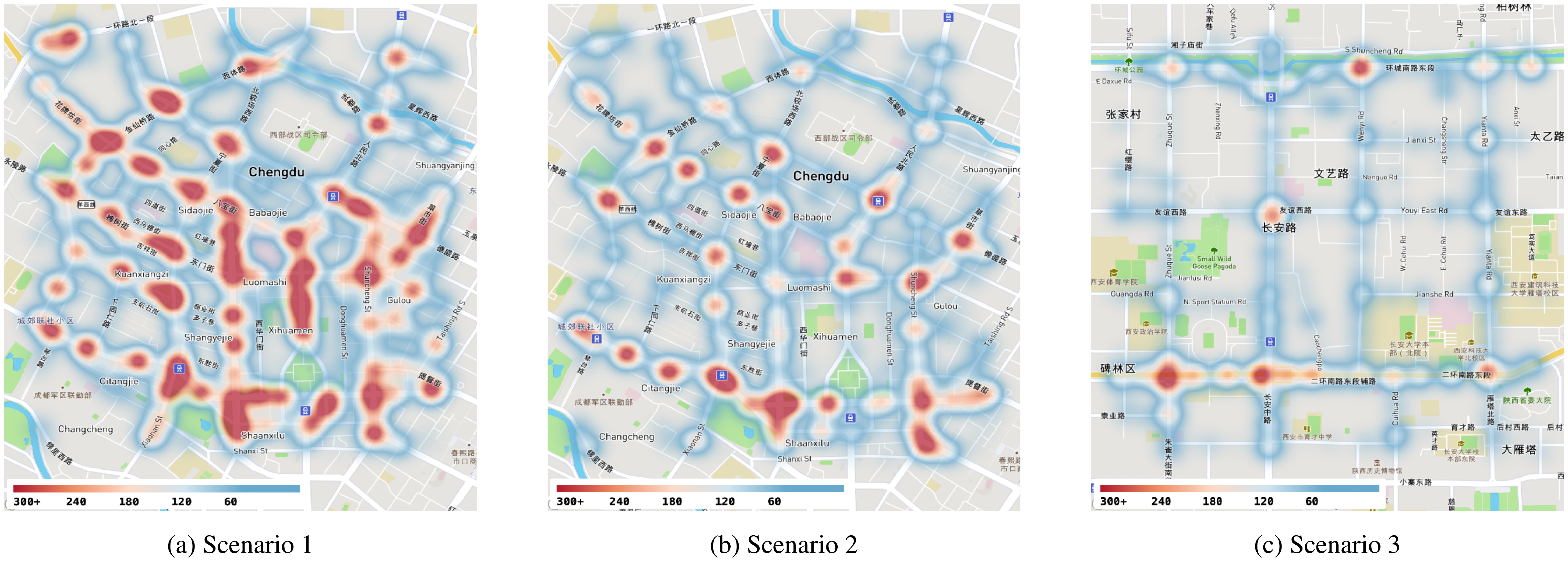}
  \caption{Heat map of the distribution of vehicles under different scenarios}
  \label{fig_4_heatmap}
\end{figure*}

In this section, we implement a simulation model by using Python 3.9 and PyTorch 1.11.0 to evaluate the performance of the proposed solution.
The simulation model \footnote{The code of the simulation model can be found at https://github.com/neardws/Multi-Agent-Deep-Reinforcement-Learning} is based on a Ubuntu 20.04 server with an AMD Ryzen 9 5950X 16-core processor @ 3.4 GHz, two NVIDIA GeForce RTX 3090 graphic processing units, and 64 GB memory.
In particular, we have examined three traffic scenarios using real-world vehicle trajectories \footnote{The  code of vehicle trajectory processing can be found at https://github.com/neardws/Vehicular-Trajectories-Processing-for-Didi-Open-Data} collected from Didi GAIA open data set\cite{didi}, including 1): a 3 km $\times$ 3 km area of Qingyang District, Chengdu, China, from 8:00 to 8:05, on 16 Nov. 2016; 2): the same area from 23:00 to 23:05, on 16 Nov. 2016; 3): a 3 km $\times$ 3 km area of Beilin District, Xian, China, from 8:00 to 8:05, on 27 Nov. 2016.
Detailed statistics including the total number of vehicle traces, average dwell time (ADT) of vehicles, the variance of dwell time (VDT), the average number of vehicles (ANV) in each second, the variance of the number of vehicles (VNV), the average speed of vehicles (ASV), and the variance of speeds of vehicles (VSV) are summarized in Table  \ref{table_traffic_characteristics}.
Fig. \ref{fig_4_heatmap} shows the heat maps of vehicle distribution within the scheduling period $T$ to better exhibit the traffic characteristic under different scenarios.
Comparing Figs. \ref{fig_4_heatmap}(a), \ref{fig_4_heatmap}(b), and \ref{fig_4_heatmap}(c), it is noted that the vehicle density in the rush hour on the weekday (i.e., around 8:00 on Nov. 16, 2016, Wed.) is much higher than that during the night (i.e., around 23:00 on Nov. 16, 2016) in the same area.
It is also much higher than that in the rush hour on the weekend (i.e., around 8:00 on Nov. 27, 2016, Sun.). 
Also, it is observed that the vehicle distribution is totally different in Fig.\ref{fig_4_heatmap}(c), which is extracted from another city.

The parameter settings are described as follows.
The data sizes of information are uniformly distributed in the range of [100B, 1MB].
The transmission power of each vehicle is set to 1 mW.
The additive white Gaussian noise and the path loss exponent of V2I communications are set to -90 dBm and 3, respectively \cite{sadek2009distributed}.
The channel fading gains of V2I communications follow the Gaussian distribution with a mean of 2 and a variance of 0.4, and the bandwidth of the edge node is set to 3 MHz \cite{wang2019delay}.
The noise uncertainty is set to follow uniformly distributed in the intervals [0, 3] dB \cite{tandra2008snr}. 

\begin{table*}[ht]\small
\centering
\caption{Traffic characteristics of each scenario}
\begin{tabular}[t]{ccccccccccc}
\hline
\hline
Scenario&Map&Number of Traces&Time&Date&ADT&VDT&ANV&VNV&ASV&VSV\\
\hline
1&Chengdu&718&08:00-08:05&Nov. 16, 2016&198.3(s)&123.8&474.6&11.6&5.22(m/s)&2.61\\
2&Chengdu&359&23:00-23:05&Nov. 16, 2016&173.7(s)&124.1&207.9&3.93&7.30(m/s)&3.16\\
3&Xian&206&08:00-08:05&Nov. 27, 2016&145.5(s)&114.7&99.9&7.65&8.06(m/s)&3.70\\
\hline
\hline
\end{tabular}
\label{table_traffic_characteristics}
\end{table*}

For the implementation of the proposed solution, the architectures and hyperparameters of the actor and critic networks are described as follows.
The local actor network is a four-layer fully connected neural network with two hidden layers, where the numbers of neurons are 64, and 32, respectively.
The architecture of the target actor network is the same as the local actor network.
The local critic network is a four-layer fully connected neural network with two hidden layers, where the numbers of neurons are 128, and 64, respectively.
The architecture of the target critic network is the same as the local critic network.
The Rectified Linear Unit (ReLU) is utilized as the activation function and the Adam optimizer is used to update network weights with a learning rate of 0.001, and the discount factor is set to 0.996.
The size of the experience replay buffer $|\mathcal{B}|$ is set to 100000, and the size of minibatch $M$ is set to 512. 
Four comparable algorithms are implemented as follows.
\begin{itemize}
	\item RA: it randomly selects one action on determining the sensing frequencies, uploading priorities, and V2I bandwidth allocation.
	\item C-DDPG \cite{mlika2022deep}: it implements an agent at the edge node to determine the sensing frequencies, uploading priorities, and V2I bandwidth allocation in a centralized way based on the system state. Meanwhile, the system reward is received by the agent to evaluate its contribution.
	\item MAC \cite{he2021efficient}: it implements agents in vehicles to decide the sensing frequencies and uploading priorities based on local observation of the physical environment, and an agent in the edge node to decide the V2I bandwidth allocation. The system reward is received by each agent to evaluate their contributions, which is the same for each agent.
	\item MAC-VBA: To enable MAC to better allocate V2I bandwidth, we further design a variant called MAC-VBA, where edge nodes allocates V2I bandwidth based on predicted vehicle trajectories and view requirements.
\end{itemize}

Furthermore, the following metrics are designed for performance evaluation.
\begin{itemize}
	\item \textit{Cumulative Reward} (CR): it is the cumulative system reward during the scheduling period $T$, which is computed by $ \sum_{\forall t \in T} r\left(\boldsymbol{a}_{S}^{t},\boldsymbol{a}_{e}^{t} \mid \boldsymbol{o}^{t}\right)$.
	\item \textit{Composition of Average Reward} (CAR): it is defined as the percentage of the normalized timeliness, completeness, and consistency in the average reward and formulated by $<\frac{3}{10}(1-\hat{\Xi}_{v}),\frac{4}{10}(1-\hat{\Phi}_{v}), \frac{3}{10}(1-\hat{\Psi}_{v})>$.
	\item \textit{Average Queuing Time} (AQT): it is defined as the sum of queuing time of the sensed information divided by the number of information during the scheduling period $T$, which is computed by $ \sum_{\forall t \in T}\{\{ \sum_{s \in S} \{\sum_{\forall d \subseteq D_{s}^t} \operatorname{q}_{d, s}^t\} / |D_{s}^t| \} / |S| \} / |T|$.
	\item \textit{Service Ratio} (SR): it is defined as the number of views, which satisfy the completeness requirement, over the total number of required views during the scheduling period $T$, which is computed by $\sum_{\forall t \in T}\sum_{\forall v \in V_e^t} \mathds{1}\{\Phi_{v} \geq \Phi_{threshold}\}  / \sum_{\forall t \in T} |V_e^t|$, and $\Phi_{threshold}$ is the completeness threshold.
\end{itemize}

\subsection{Results and Analysis}
\begin{figure}
\centering
  \includegraphics[width=0.99\columnwidth]{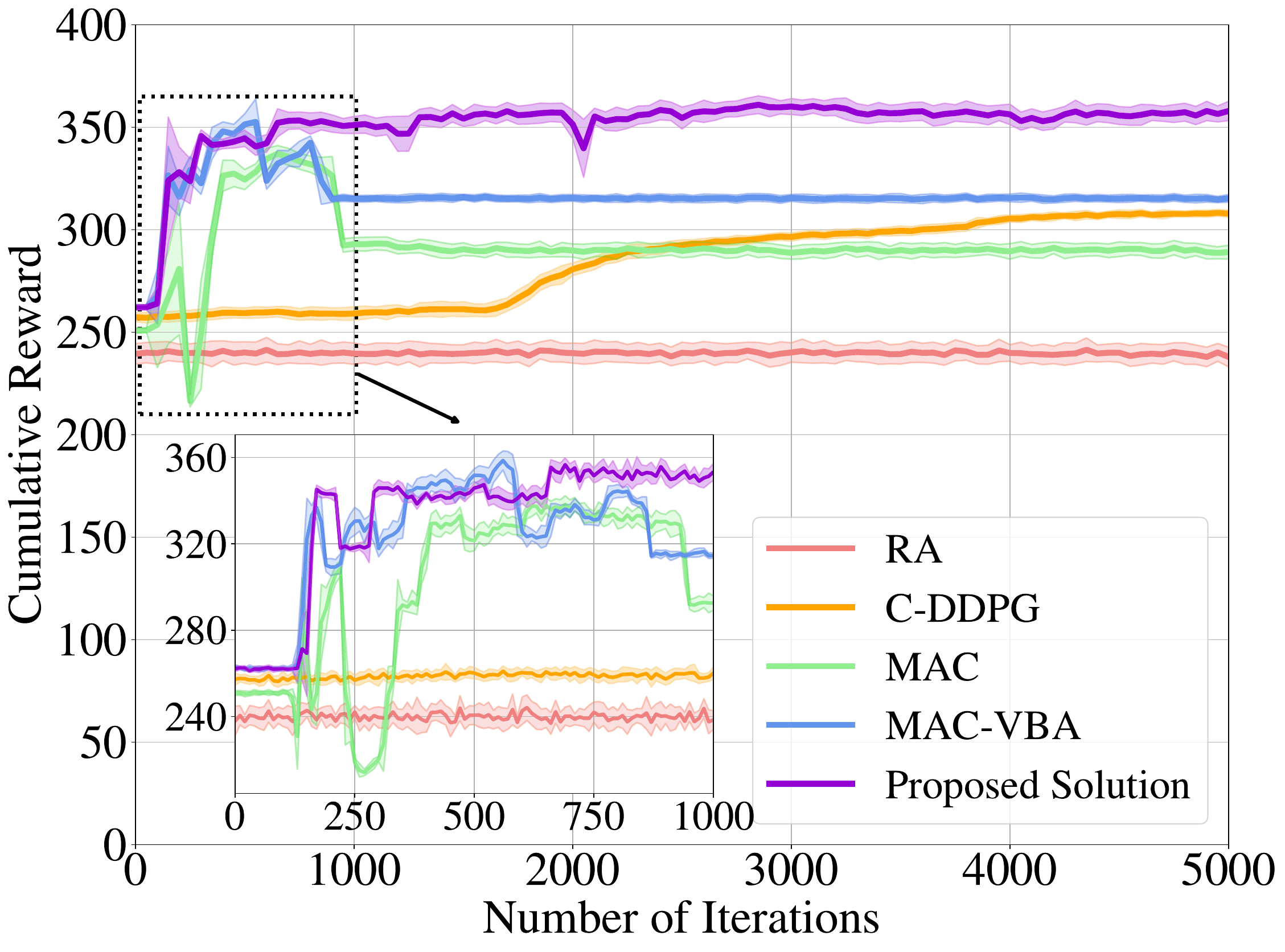}
  \caption{Convergence comparison}
  \label{fig_4_convergence}
\end{figure} 

\textit{1) Algorithm Convergence:}
Fig. \ref{fig_4_convergence} compares the CR of the five algorithms in terms of convergence speed.
As noted, the proposed solution converges the fastest (around 660 iterations) and achieves the highest CR value (around 357). 
In contrast, C-DDPG, MAC and MAC-VBA converge after around 4500, 950, and 870 iterations and achieve the CR around 307, 290 and 315, respectively.
RA achieves a CR of around 241 as the baseline. 
We note that the proposed solution achieves about 16.3\%, 23.1\%, and 13.3\% improvement with respect to CR and about 6.8$\times$, 1.4$\times$, and 1.3$\times$ improvement with respect to converge speed compared with C-DDPG, MAC, and MAC-VBA, respectively.
The primary reason is that the proposed solution is designed to maintain a stable communication environment for vehicles, which makes the training of actor and critic networks in vehicles more efficient. 
On the other hand, due to the smaller action space of the proposed solution model, the proposed solution converges much faster than C-DDPG, which decides the actions on determining the sensing frequencies, uploading priorities, and V2I bandwidth allocation simultaneously using a DDPG agent. 

\begin{figure*}[t!]
  \centering
  \includegraphics[height=4.12in]{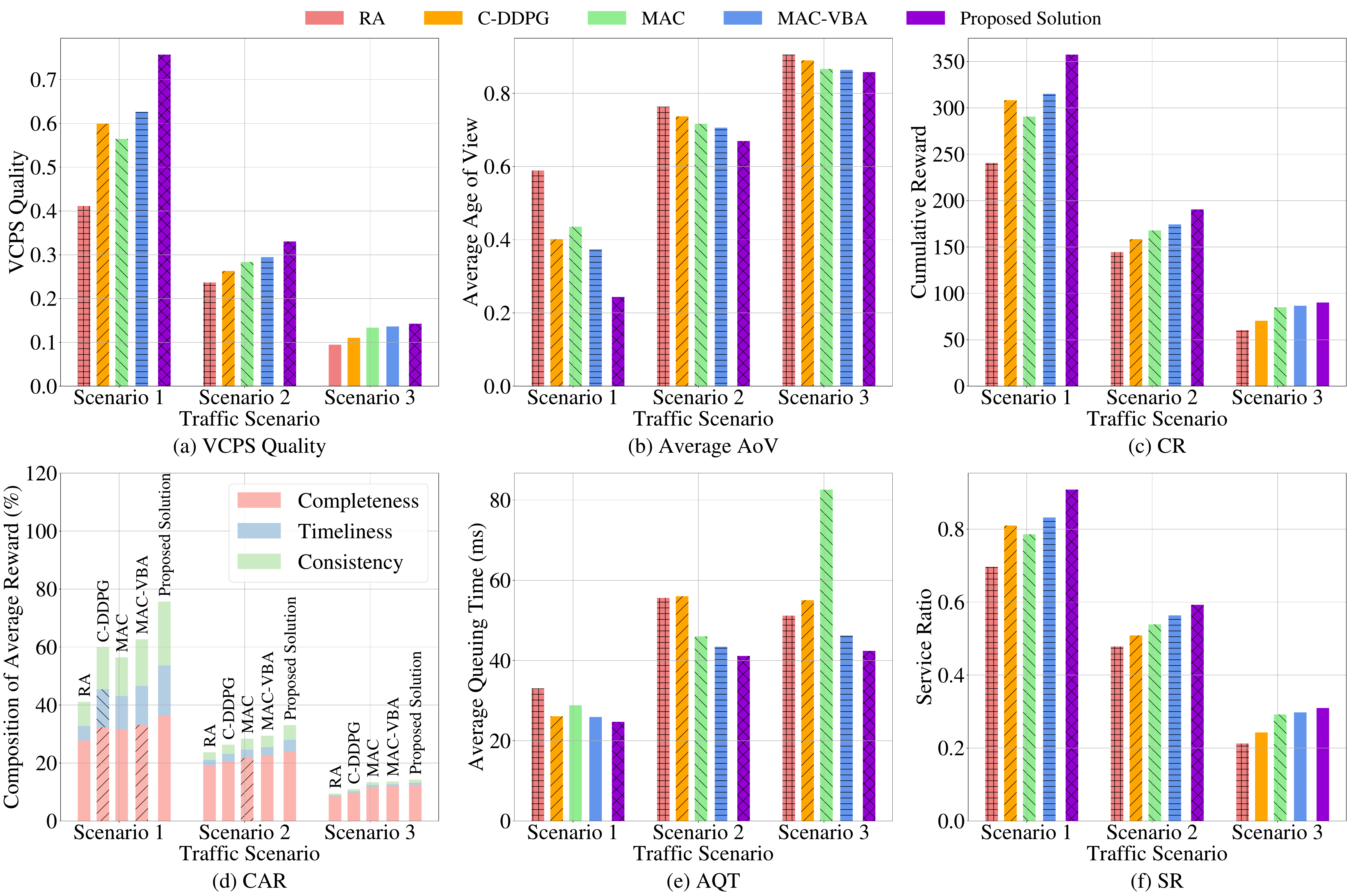}
  \caption{Performance comparison under different traffic scenarios}
  \label{fig_5_scenarios}
\end{figure*}

\textit{2) Effect of traffic scenarios:}
Fig. \ref{fig_5_scenarios} compares the performance of the five algorithms under different traffic scenarios.
Fig. \ref{fig_5_scenarios}(a) compares the VCPS quality of the five algorithms.
As demonstrated, the proposed solution achieves the highest VCPS quality under all scenarios.
In particular, the proposed solution improves the VCPS quality by 58.0\%, 27.1\%, 19.1\%, and 12.5\%  on average over RA, C-DDPG, MAC, and MAC-VBA, respectively, under different traffic scenarios.
Fig. \ref{fig_5_scenarios}(b) shows the average AoV of the five algorithms.
It is expected that the proposed solution achieves the lowest average AoV under all the scenarios.
Fig. \ref{fig_5_scenarios}(c) compares the CR of the five algorithms.
As noted, the CR of the proposed solution is higher than RA, C-DDPG, MAC, and MAC-VBA.
Meanwhile, the CR of the proposed solution and MAC-VBA are similar under scenario 3.
The reason is that the lower vehicle density and higher traffic dynamic in scenario 3 make the data uploading more difficult than that in scenarios 1 and 2.

Fig. \ref{fig_5_scenarios}(d) breaks down the average reward into three parts, which demonstrates the proportion of timeliness, completeness, and consistency, respectively.
It is observed that the timeliness and consistency of the five algorithms are very small under scenario 3.
This is mainly because the requirements of timeliness and consistency can hardly be satisfied when the view is incomplete.
Figs. \ref{fig_5_scenarios}(e) and \ref{fig_5_scenarios}(f) compare the AQT and SR of the five algorithms under different traffic scenarios.
It demonstrates that the proposed solution achieves the lowest AQT, and maintains the highest SR under all scenarios.

\begin{figure*}[t!]
  \centering
  \includegraphics[height=4.12in]{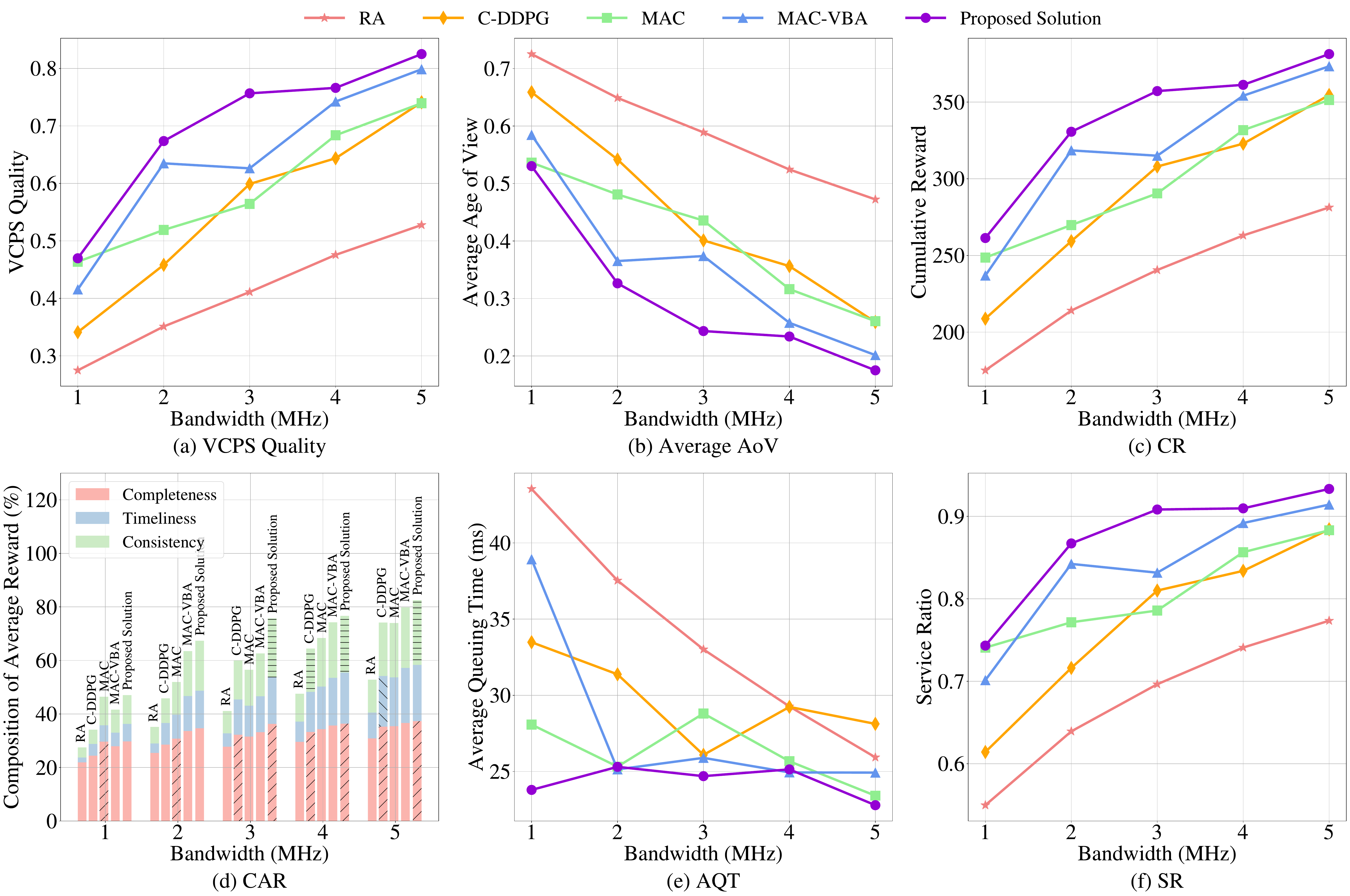}
  \caption{Performance comparison under different V2I bandwidths}
  \label{fig_5_bandwidth}
\end{figure*}

\textit{3) Effect of V2I bandwidths:}
Fig. \ref{fig_5_bandwidth} compares the performance of the five algorithms under different V2I bandwidths.
In this set of experiments, we consider the V2I bandwidth of edge nodes increases from 1 MHz to 5 MHz.
A larger bandwidth represents that more information can be uploaded via V2I communications. 
Fig. \ref{fig_5_bandwidth}(a) compares the VCPS quality of the five algorithms.
With the increasing bandwidth, the VCPS quality of all algorithms increases accordingly.
The VCPS quality of the proposed solution outperforms RA, C-DDPG, MAC, and MAC-VBA by around 72.9\%, 28.3\%, 17.8\% and 9.3\%, respectively, under different bandwidths of the edge node.
Fig. \ref{fig_5_bandwidth}(b) compares the average AoV of the five algorithms.
In particular, the proposed solution achieves the lowest average AoV across all cases.
Fig. \ref{fig_5_bandwidth}(c) compares the CR of the five algorithms. 
As expected, the performance of all five algorithms gets better when the bandwidth increases. 
Specifically, the proposed solution achieve around 75.1\%, 29.4\%, 22.7\%, and 10.6\% improvement in CR than RA, C-DDPG, MAC, and MAC-VBA, respectively.

Fig. \ref{fig_5_bandwidth}(d) compares the CAR of the five algorithms.
The proposed solution achieves a better performance than the other four algorithms, particularly, in terms of timeliness and consistency of average reward.
This is because the cooperation of sensing and uploading information among vehicles is more efficient in the proposed solution under the limited bandwidth.
Fig. \ref{fig_5_bandwidth}(e) compares the AQT of the five algorithms. 
As noted, the AQT of the proposed solution maintains the lowest under different edge bandwidths, which reflects that the designed proposed solution can allocate the bandwidth more efficiently. 
The advantage can be further justified by Fig. \ref{fig_5_bandwidth}(f), which shows the SR of the five algorithms.
The SR of the proposed solution remains at the highest level across all cases.

\begin{figure*}[t!]
  \centering
  \includegraphics[height=4.12in]{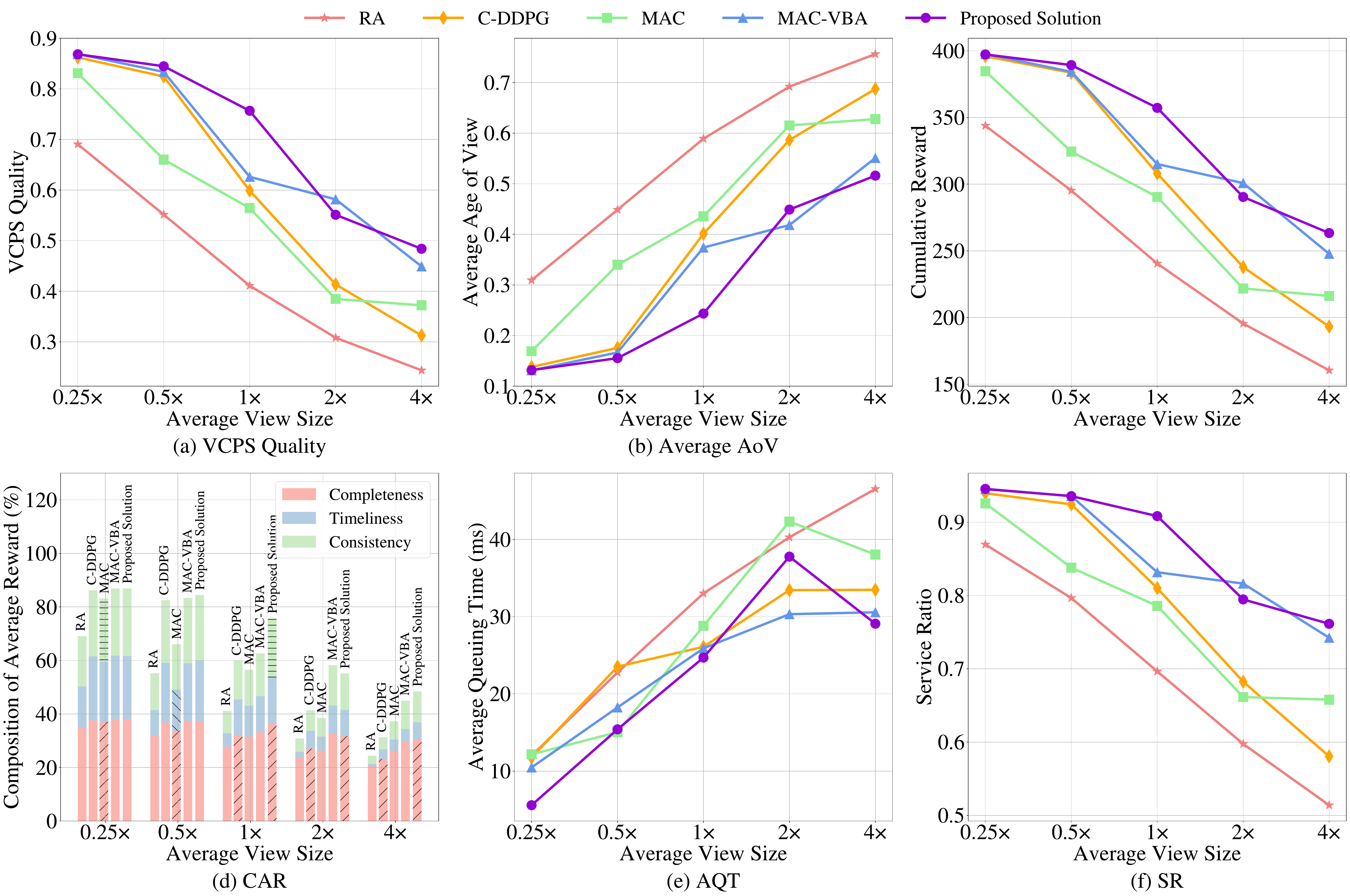}
  \caption{Performance comparison under different requirements on views}
  \label{fig_6_datasize}
\end{figure*}

\textit{4) Effect of view requirements:}
Fig. \ref{fig_6_datasize} compares the performance of the five algorithms under different requirements on views, in which the average size of views required by ITS applications increases from 0.25$\times$ to 4$\times$, and the average view size of 1$\times$ is around 6.46 MB.
Fig. \ref{fig_6_datasize}(a) compares the VCPS quality of the five algorithms.
As expected, the performance of all five algorithms gets worse when the average view size increases.
The proposed solution outperforms RA, C-DDPG, MAC, and MAC-VBA by around 68.1\%, 23.5\%, 27.9\% and 4.9\%, respectively, in terms of maximizing the VCPS quality under different application requirements on views.
Figs. \ref{fig_6_datasize}(b) and \ref{fig_6_datasize}(c) compare the average AoV and CR of the five algorithms.
When the average view size is small (i.e., around 1.62 MB), the average AoV in the proposed solution is slightly lower than that in MAC and MAC-VBA.
Meanwhile, note that the CR of the proposed solution, MAC, and MAC-VBA are similar. 
The reason is that a smaller data size has a higher probability of being successfully uploaded. 

Fig. \ref{fig_6_datasize}(d) compares the CAR of the five algorithms. 
It is observed that the performance difference between the proposed solution and MAC-VBA is small when the average view size increases from 0.25$\times$ to 0.5$\times$.
The reason is the scheduling effect is not significant when there are sufficient resources to meet the requirements of a smaller average view size (i.e., around 1.62 MB and 3.23 MB).
Figs. \ref{fig_6_datasize}(e) and \ref{fig_6_datasize}(f) compare the AQT and SR of the five algorithms, showing that the proposed solution can remain the lowest AQT, and meanwhile achieve the highest SR in most cases.
It is noted that MAC-VBA achieves the lowest AQT and the highest SR when the average view size is 2×, which reflects that the proposed VBA scheme can allocate the bandwidth more efficiently.

\section{Conclusion and Future Work}
In this paper, a new metric AoV was designed to evaluate the quality of the logical view constructed at the edge in terms of timeliness, completeness, and consistency of heterogeneous information in VCPS. On this basis, the problem was formulated to maximize the VCPS quality. Further, a tailored solution based on multi-agent DRL was proposed, in which vehicles act as independent agents to determine the sensing frequencies and uploading priorities. Edge nodes allocated the V2I bandwidth based on VBA scheme by considering vehicle trajectories and view requirements. The DR-based credit assignment scheme was adopted to divide the system reward based on the difference between the system reward and achieved reward without the action of vehicle, which are utilized to evaluate the individual contributions of vehicles. Finally, a comprehensive performance evaluation was conducted to demonstrate the significance of the newly designed metric AoV and the superiority of the proposed solution. In particular, the proposed solution outperforms RA, C-DDPG, MAC, and MAC-VBA by around 61.8\%, 23.8\%, 22.0\%, and 8.0\%, respectively, in terms of maximizing the VCPS quality. Meanwhile, compared with C-DDPG, MAC, and MAC-VBA, the proposed solution speeds up the convergence by around 6.8$\times$, 1.4$\times$, and 1.3$\times$, respectively.

As an early stage of exploring cooperative sensing and information fusion in VCPS, this work focused on the AoV modeling and evaluation within the coverage of a single edge node. In our future work, the cooperation among the edge nodes will be further investigated to extend the supported ITS applications as well as enhance overall system  performance. Also, the transmission power control and allocation will be further investigated when considering the overall power consumption and spatial reusability of V2V communication.

 
\bibliographystyle{IEEEtran} 
\bibliography{reference}



\vspace{-20pt}
\begin{IEEEbiography}[{\includegraphics[width=1in,height=1.25in,clip,keepaspectratio]{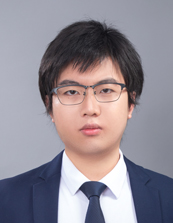}}]{Xincao Xu}
received the B.S. degree in network engineering from the North University of China, Taiyuan, China, in 2017. He is currently pursuing the Ph.D. degree in computer science at Chongqing University, Chongqing, China. His research interests include vehicular networks, edge computing, and deep reinforcement learning.
\end{IEEEbiography}

\vspace{-20pt}

\begin{IEEEbiography}[{\includegraphics[width=1in,height=1.25in,clip,keepaspectratio]{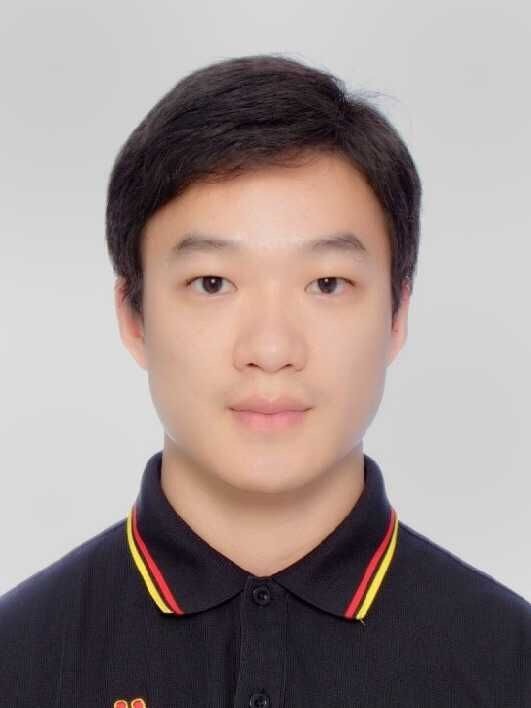}}]{Kai Liu}
(Senior Member, IEEE) received the Ph.D. degree in computer science from the City University of Hong Kong in 2011. He is currently a Full Professor with the College of Computer Science, Chongqing University, China. From 2010 to 2011, he was a Visiting Scholar with the Department of Computer Science, University of Virginia, Charlottesville, VA, USA. From 2011 to 2014, he was a Postdoctoral Fellow with Nanyang Technological University, Singapore, City University of Hong Kong, and Hong Kong Baptist University, Hong Kong. His research interests include mobile computing, pervasive computing, intelligent transportation systems, and the Internet of Vehicles.
\end{IEEEbiography}

\begin{IEEEbiography}[{\includegraphics[width=1in,height=1.25in,clip,keepaspectratio]{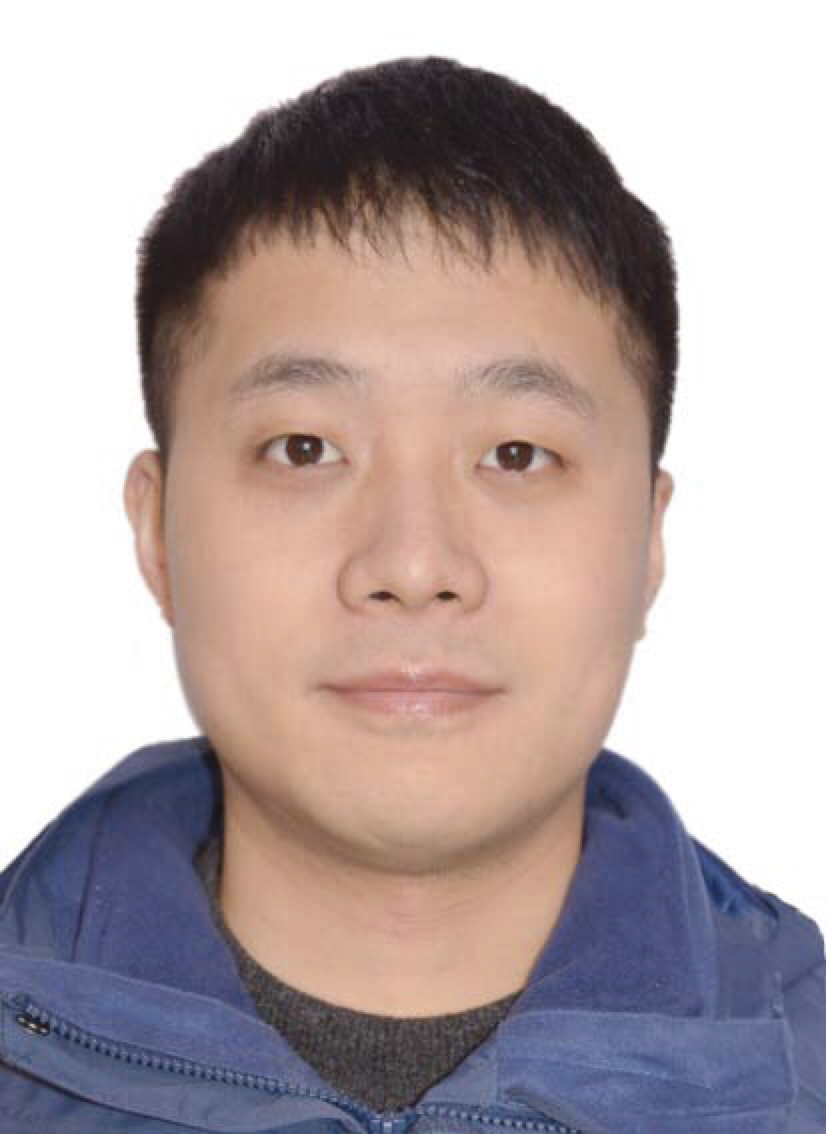}}]{Penglin Dai}
(Member, IEEE) received the B.S. degree in mathematics and applied mathematics and the Ph.D. degree in computer science from Chongqing University, Chongqing, China, in 2012 and 2017, respectively. He is currently an Associate Professor with the School of Information Science and Technology, Southwest Jiaotong University, Chengdu, China. His research interests include intelligent transportation systems and vehicular cyber-physical systems.
\end{IEEEbiography}

\begin{IEEEbiography}[{\includegraphics[width=1in,height=1.25in,clip,keepaspectratio]{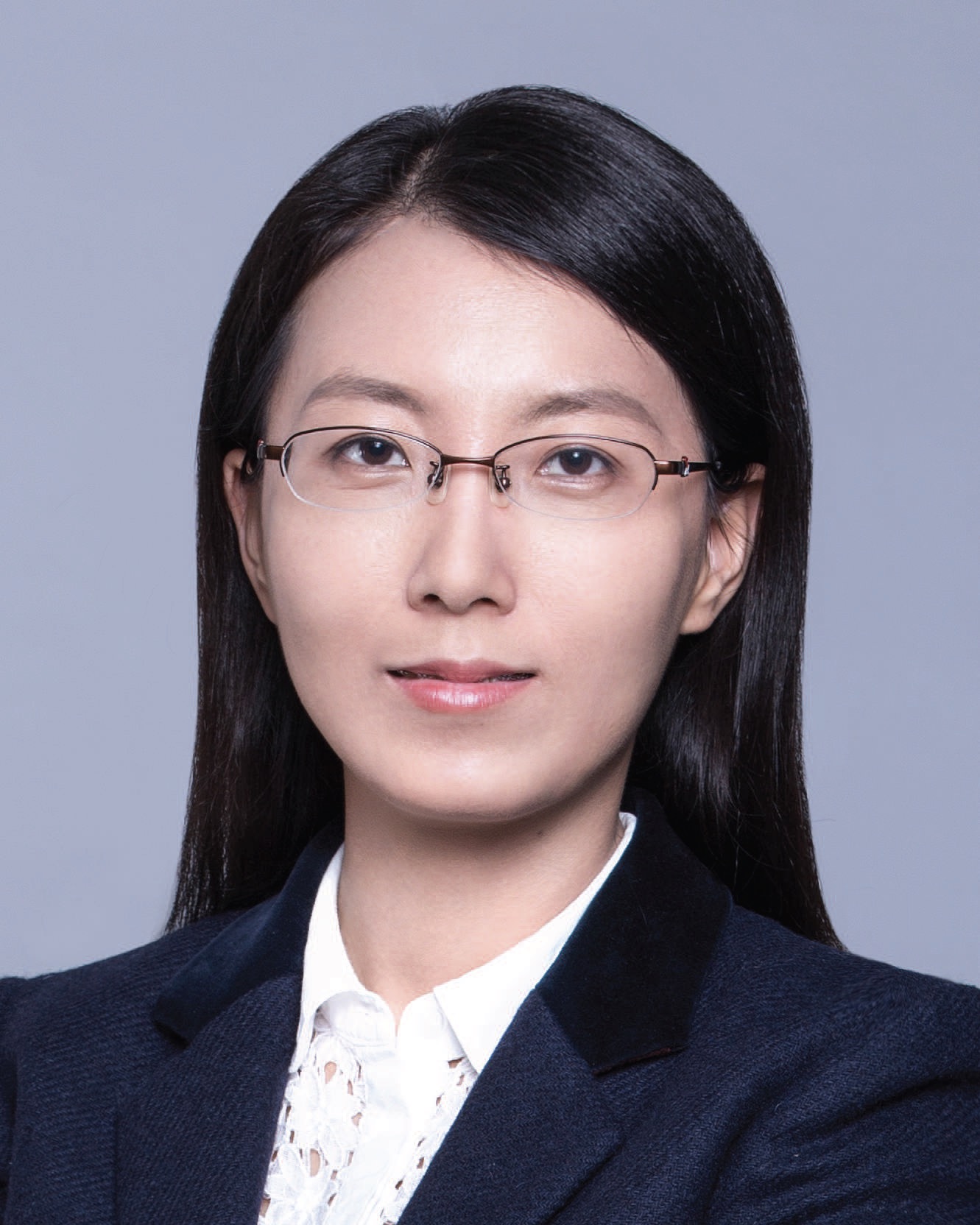}}]{Ruitao Xie}
received the Ph.D. degree in computer science from City University of Hong Kong in 2014, and the B.Eng. degree from Beijing University of Posts and Telecommunications in 2008. She is currently an Assistant Professor with the College of Computer Science and Software Engineering, Shenzhen University, Shenzhen, China. Her research interests include AI networking and mobile computing, distributed systems, and cloud computing.
\end{IEEEbiography}

\begin{IEEEbiography}[{\includegraphics[width=1in,height=1.25in,clip,keepaspectratio]{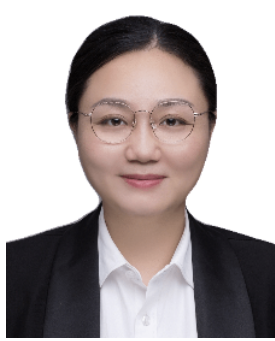}}]{Jingjing Cao}
received her Ph.D. degree in Computer Sciences from City University of Hong Kong, Hongkong, China, in 2013. She is currently an Association Professor of the School of Transportation and Logistics Engineering, Wuhan University of Technology, Hubei, China. Her current research interests include machine learning, pervasive computing and their applications in transportation and logistics.
\end{IEEEbiography}

\begin{IEEEbiography}[{\includegraphics[width=1in,height=1.25in,clip,keepaspectratio]{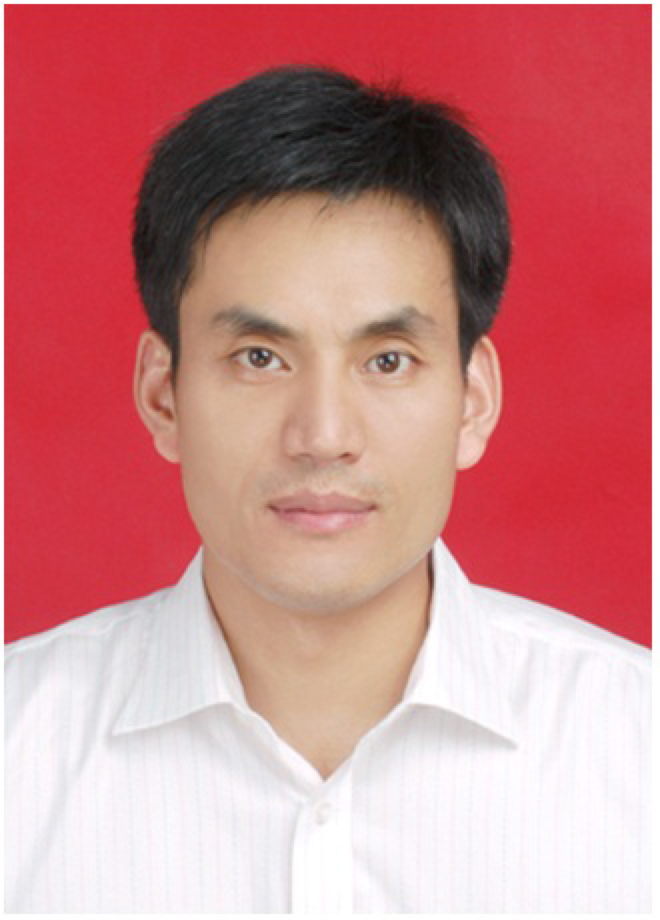}}]{Jiangtao Luo}
(Senior Member, IEEE) received the B.S. degree from Nankai University in 1993 and the Ph.D. degree from the Chinese Academy of Science in 1998. He is currently a Full Professor, a Ph.D. Supervisor, and the Deputy Dean of the Electronic Information and Networking Research Institute, Chongqing University of Posts and Telecommunications (CQUPT). He has been visiting scholars with the University of Hamburg, Germany, and the University of Southern California, Los Angeles, CA, USA, from 2015 to 2016. His research interests are network data analysis, visual big data, and future Internet architecture. He has published more than 100 articles and owned 30 patents in these fields. He was awarded the Chinese State Award of Scientific and Technological Progress in 2011, the Chongqing Provincial Award of Scientific and Technological Progress twice in 2010 and 2007, respectively, and the Chongqing Science and Technology Award for Youth in 2010.

\end{IEEEbiography}

\vfill

\end{document}